\documentclass[prl,twocolumn,superscriptaddress,amsmath,amssymb,floatfix,preprintnumbers]{revtex4-1}
\usepackage{amsmath}
\usepackage{graphicx}
\usepackage{dcolumn}
\usepackage{color}
\usepackage{mathrsfs}
\usepackage{physics}
\usepackage{float}
\usepackage{soul}
\usepackage{xr}

\setstcolor{black} 

\makeatletter
\newcommand*{\addFileDependency}[1]{
  \typeout{(#1)}
  \@addtofilelist{#1}
  \IfFileExists{#1}{}{\typeout{No file #1.}}
}
\makeatother
 
\newcommand*{\myexternaldocument}[1]{%
    \externaldocument{#1}%
    \addFileDependency{#1.tex}%
    \addFileDependency{#1.aux}%
}

\myexternaldocument{si}

\begin{document}

\title{Strain Engineering of Altermagnetic Symmetry in Epitaxial RuO$_2$ Films}

\date{ \today }

\author{Johnathas D. S. Forte}\thanks{These authors contributed equally to this work.}
\affiliation{Department of Electrical and Computer Engineering, University of Minnesota, Minneapolis, Minnesota 55455, USA}

\author{Seung Gyo Jeong}\thanks{These authors contributed equally to this work.}
\affiliation{Department of Chemical Engineering and Materials Science, University of Minnesota-Twin Cities, Minneapolis, Minnesota 55455, USA}

\author{Anand Santhosh}
\affiliation{Department of Chemical Engineering and Materials Science, University of Minnesota-Twin Cities, Minneapolis, Minnesota 55455, USA}

\author{Seungjun Lee}\email{sjunlee@khu.ac.kr}
\affiliation{Department of Electrical and Computer Engineering, University of Minnesota, Minneapolis, Minnesota 55455, USA}
\affiliation{Department of Applied Physics, Kyung Hee University, Yongin 17104, Republic of Korea}

\author{Bharat Jalan}\email{bjalan@umn.edu}
\affiliation{Department of Chemical Engineering and Materials Science, University of Minnesota-Twin Cities, Minneapolis, Minnesota 55455, USA}

\author{Tony Low}\email{tlow@umn.edu}
\affiliation{Department of Electrical and Computer Engineering, University of Minnesota, Minneapolis, Minnesota 55455 - USA}
\affiliation{School of Physics and Astronomy, University of Minnesota, Minneapolis, Minnesota 55455, USA}

\begin{abstract}
The magnetic ground state of RuO$_2$ has been under intense debate. Using first-principles calculations, we show that compressive strain along [001] direction stabilizes an altermagnetic phase in RuO$_2$ thin films grown on (100) and (110) TiO$_2$ substrates.
We further identify that compressive strain enhances the density of states near the Fermi level, resulting in a Fermi surface instability and the emergence of altermagnetism.
The magnitude of strain and the associated increase in the density of states can be tuned by varying the film thickness, as systematically confirmed by x-ray diffraction and photoemission spectroscopy measurements.
Symmetry analysis further reveals that (100) RuO$_2$ hosts an ideal altermagnetic order, whereas broken symmetry in (110) films leads to an uncompensated ferrimagnetic state.  Finally, we discuss the effects of Hubbard $U$ parameters and evaluate the realistic tunneling magnetoresistance of (100) RuO$_2$.
\end{abstract}

\maketitle

\paragraph*{Introduction} Altermagnetism, a recently identified collinear magnetic phase, exhibits compensated local magnetic moments while breaking time-reversal ($\mathcal{T}$) symmetry, leading to non-relativistic spin-momentum locking.~\cite{10.1126/sciadv.aaz8809,PhysRevX.12.040501,vsmejkal2022beyond,yuan2021prediction}
This unique property opens a novel possibility for energy-efficient and ultrafast spintronic applications, potentially overcoming the fundamental limitations of conventional ferromagnetic and antiferromagnetic spintronics.~\cite{shao2021spin,vsmejkal2023chiral,bai2024altermagnetism}
After identifying the first altermagnetic material, RuO$_2$,~\cite{berlijn2017itinerant,10.1126/sciadv.aaz8809} significant research efforts have been dedicated to discovering novel altermagnetic materials.~\cite{krempasky2024altermagnetic,lee2024broken,zhou2025manipulation,han2024electrical,takagi2025spontaneous,Jiang2025,Zhang2025}

Despite these advancements, the magnetic ground state of RuO$_2$ still remains under intense debate. Early experimental studies using neutron diffraction~\cite{berlijn2017itinerant} and resonant x-ray scattering~\cite{zhu2019anomalous} reported antiferromagnetic order in RuO$_2$, and density functional theory (DFT) calculations with the Hubbard $U$ correction revealed its altermagnetic properties.~\cite{10.1126/sciadv.aaz8809,PhysRevX.12.040501} 
These reports attracted explosive attention, and following experiments revealed signatures consistent with altermagnetic order.~\cite{feng2022anomalous,tschirner2023saturation,bose2022tilted,bai2022observation,liao2024separation,weber2024all,fedchenko2024observation}
However, more recently, conflicting evidence has emerged from \textcolor{black}{x-ray diffraction (XRD),} muon spin rotation and neutron diffraction experiments, which reported a nonmagnetic ground state in bulk RuO$_2$ and/or in relatively thick films.~\cite{hiraishi2024nonmagnetic,kessler2024absence,occhialini2025structural,kiefer2025crystal}
Subsequent experiment and theoretical results have also claimed its nonmagnetic ground state, further fueling the debate.~\cite{liu2024absence,peng2024universal,wenzel2025fermi,PhysRevB.109.134424,wickramaratne2025effects}

\textcolor{black}{{It is notable that, when RuO$_2$ approaches the atomically thin limit, its altermagnetic behavior becomes more consistently observable, suggesting that both sample thickness and epitaxial strain play important roles in its magnetism. For instance, mirror-even spin splitting have been observed in spin- and angle-resolved photoemission spectroscopy in a 2 nm (110) RuO$_2$ thin film.~\cite{yichen2025} Furthermore, second-harmonic generation measurements have revealed time-reversal symmetry breaking in ultrathin ($<$4~nm) (110) RuO$_2$ films epitaxially grown on (110) TiO$_2$, which is absent in bulk single crystal RuO$_2$, further underscoring the critical influence of epitaxial strain~\cite{jeong2024altermagnetic}. 
The observation of a sizable anomalous Hall effect in the ultrathin films consistently suggests a strain-driven stabilization of altermagnetic order~\cite{jeong2025metallicity}. Nevertheless, the fundamental physical relationship between altermagnetic order and the lattice parameters of RuO$_2$ remains largely unexplored.}}

\begin{figure*}[t]
    \centering
    \includegraphics[width=\linewidth]{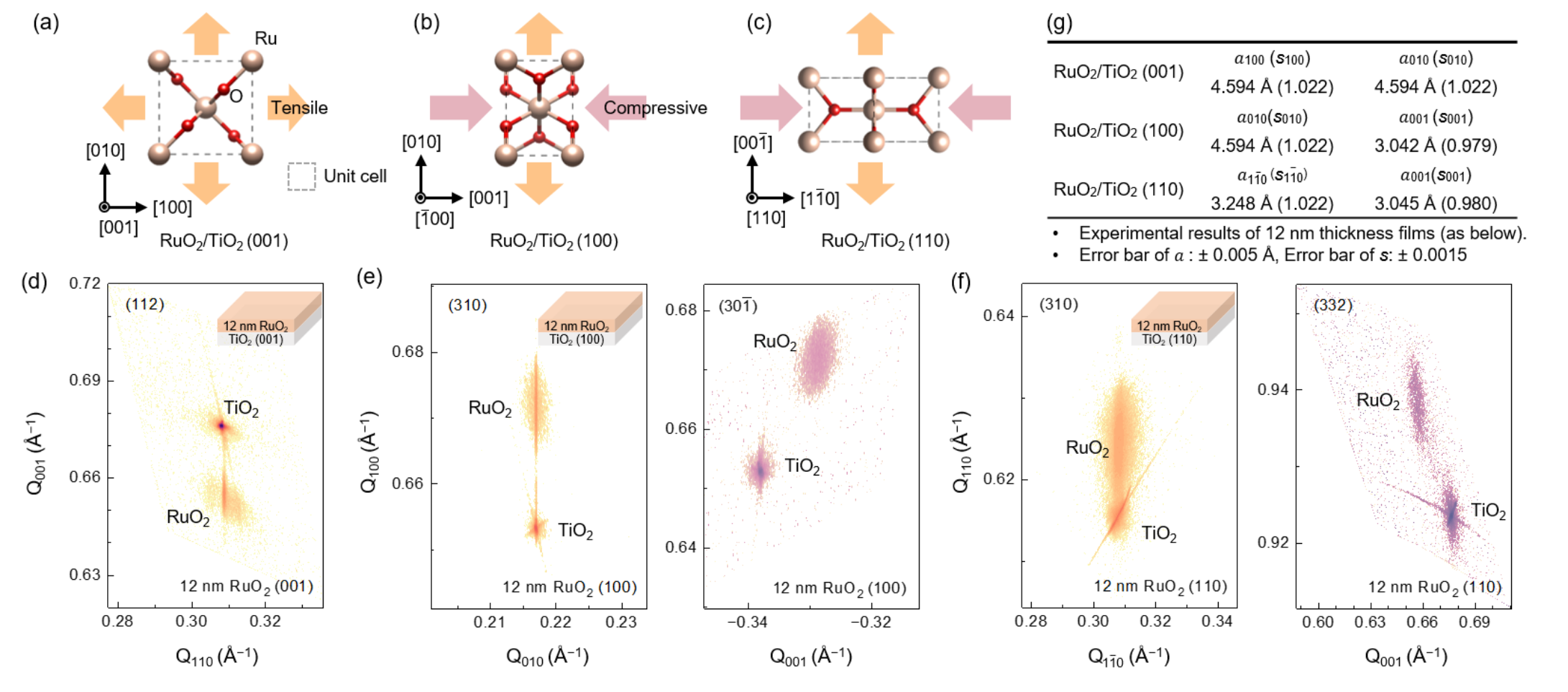}
 \caption{(a-c) Schematic illustrations of epitaxial strain in RuO\textsubscript{2} films grown on TiO\textsubscript{2} substrates with three orientations: (001), (100), and (110). Orange and pink arrows represent tensile and compressive strain components, respectively. (d-f) XRD RSMs of 12~nm thick RuO\textsubscript{2} films grown on (001), (100), and (110) TiO\textsubscript{2} substrates. (g) Summary of lattice constants \textcolor{black}{(with respect to the primitive unit cell)} and calculated strain factors (\textit{s}) for each crystallographic orientation of RuO\textsubscript{2}/TiO\textsubscript{2} heterostructures.}
    \label{fig:Fig0}
\end{figure*}

Here, we systematically investigate the strain effects in RuO$_2$ films grown on TiO$_2$ substrates and their influence on magnetic order. The \textcolor{black}{XRD} measurements reveal that the lattice constants of RuO$_2$ strongly depend on film thickness due to epitaxial strain from the TiO$_2$ substrate.
Our first-principles calculations show that compressive strain along the [001] direction enhances the density of states near the Fermi level, driving Fermi surface instabilities and stabilizing the altermagnetic phase. Notably, this strain-induced magnetism emerges in RuO$_2$ films on (100) and (110) oriented TiO$_2$ without requiring Hubbard $U$ corrections or additional hole doping. Furthermore, we find that additional symmetry breaking in (110) RuO$_2$ induces an uncompensated ferrimagnetic order, in contrast to the ideal altermagnetic phase in (100) RuO$_2$. Finally, we analyze spin–momentum locking and the Fermi surface of altermagnetic (100) RuO$_2$, and evaluate the tunneling magnetoresistance ratio within a realistic range of Hubbard $U$ parameters.

\paragraph*{Epitaxial strain of RuO$_2$ on TiO$_2$ substrate} The lattice constants of RuO$_2$ thin films can be different from its bulk values. In the former, the substrate imposes epitaxial strain in RuO$_2$ arising from the lattice constant mismatch, which can vary depending on the crystallographic direction of epitaxial growth.
At room temperature, bulk RuO$_2$ has tetragonal lattice constants of $a_{100}^{\rm{RuO_2}} =a_{010}^{\rm{RuO_2}}=4.492$~{\AA} and $a_{001}^{\rm{RuO_2}}=3.106$~{\AA},~\cite{PhysRevLett.118.077201} while bulk TiO$_2$ has $a_{100}^{\rm{TiO_2}}=a_{010}^{\rm{TiO_2}}=4.594$~{\AA} and $a_{001}^{\rm{TiO_2}}=2.959$~{\AA}.~\cite{doi:10.1021/ja00246a021} Here, the subscripts represent the Miller indices for the crystallographic orientations. 

For the (001) plane, as shown in Fig.~\ref{fig:Fig0}(a), the in-plane lattice mismatch is isotropic; +2.2\% along both [100] and [010] directions. 
For convenience, we define the epitaxial strain factor as $s_{hkl}$ = $a_{hkl}$/$a^{\rm{RuO_2}}_{hkl}$, where $a_{hkl}$ ($a_{hkl}^{\rm{RuO_2}}$) is the lattice constant of strained (bulk) RuO$_2$ along the direction $[hkl]$, resulting in $s_{100}=s_{010}=1.022$. 
In contrast, both (100) and (110) planes exhibits anisotropic lattice mismatch: $-4.7$\% ($s_{001}=0.953$) along [001] and +2.2\% ($s_{010}=s_{1\bar{1}0}=1.022$) along [100] or [1$\bar{1}$0], shown in Fig.~\ref{fig:Fig0}(b) and (c).
To experimentally elucidate the strain state and its crystallographic anisotropy, we grew three different 12 nm RuO$_2$ films on TiO$_2$ (001), (100), and (110) substrates using hybrid molecular beam epitaxy (MBE). XRD reciprocal space mapping (RSM) was performed around asymmetric reflections for each sample, as shown in Fig. \ref{fig:Fig0}(d)-(f). For the (112) reflection of the 12 nm RuO$_2$/TiO$_2$ (001) film (Fig. \ref{fig:Fig0}(d)), the RuO$_2$ peak is vertically aligned with the TiO$_2$ substrate peak along the in-plane (110) reciprocal space vector ($Q_{110}$). This alignment indicates that the in-plane lattice constants of RuO2 are fully locked to those of the substrate, confirming coherent epitaxial strain. In contrast, for the RuO$_2$/TiO$_2$ (100) film (Fig. \ref{fig:Fig0}(e)), the (310) and (30-1) RSMs reveal anisotropic strain relaxation. While the RuO$_2$ peak remains vertically aligned with the substrate along $Q_{010}$, indicating coherent strain along that direction, the peak shifts laterally along $Q_{001}$. A similar anisotropic behavior is observed for the RuO$_2$/TiO$_2$ (110) film (Fig. \ref{fig:Fig0}(f)). The RuO$_2$ peak remains aligned with the substrate along $Q_{1\bar{1}0}$, whereas a clear lateral shift is observed along $Q_{001}$. This crystal direction-dependent relaxation along $Q_{001}$ for both (100) and (110) cases evidence anisotropic in-plane lattice relaxation. The extracted lattice parameters and corresponding strain values are summarized in Fig. \ref{fig:Fig0}(g) and Table S1 in the Supplementary Information (SI)\cite{SI}. These experimentally quantified strain states provide a realistic structural basis for interpreting the theoretically predicted evolution of magnetic properties under anisotropic strain modulations.

\begin{figure}[t]
    \centering
    \includegraphics[width=\linewidth]{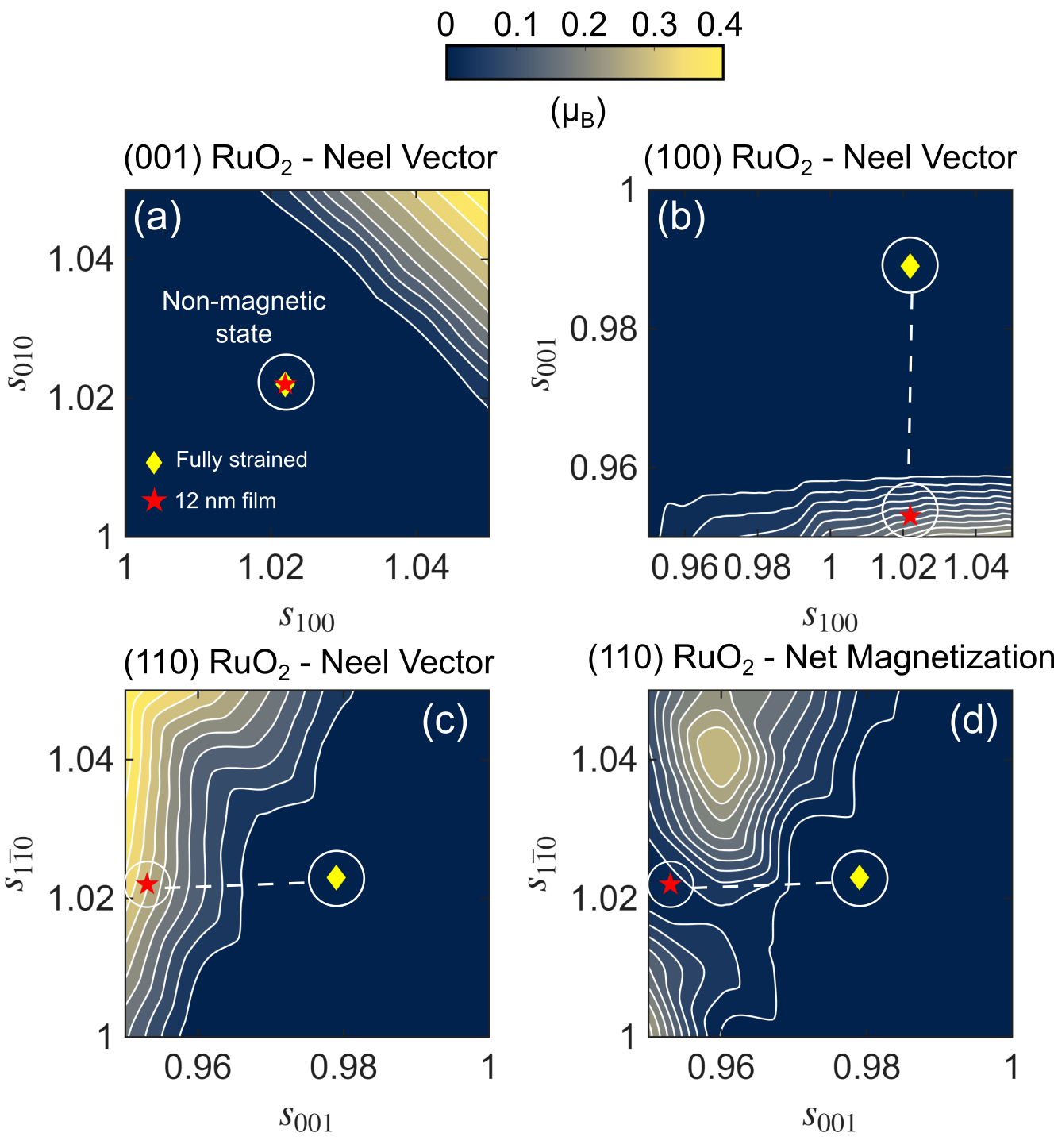}
    \caption{Strain-magnetization Neel vector phase diagrams for (a) (001), (b) (100) and (c) (110) RuO$_2$. The red stars and \textcolor{black}{yellow} diamonds in each panel indicate lattice constants of fully strained and partially strained RuO$_2$ on TiO$_2$, respectively. (d) Strain-net magnetic moments for (110) RuO$_2$.}
    \label{Fig1}
\end{figure}

\paragraph*{Strain -- magnetization phase diagrams} To clearly understand the relation between the lattice parameter and magnetic ground state of RuO$_2$, we map the strain--magnetization phase diagram of (001), (100), and (110) RuO$_2$ through first-principles DFT calculations, as shown in Fig.\ref{Fig1}(a)-(c). The phase diagrams explore a range of strain states imposed by different types of substrates. \textcolor{black}{Strain was simulated by varying the in-plane lattice constants for each case and finding the new equilibrium out-of-plane lattice constant. We highlight with red stars the points on the phase diagrams corresponding to fully strained RuO$_2$ films grown on TiO$_2$, estimated from the bulk lattice constants. By using XRD, although the fully strained lattice of a single RuO$_2$ film ($< 4$ nm) on TiO$_2$ (100) and (110) substrates could not be unambiguously resolved, fully strained ultrathin layers were observed in a 2 nm RuO$_2$/TiO$_2$ (110) superlattice \cite{jeong2024altermagnetic}. In order to compare our theoretical calculations, we also add the \textcolor{black}{yellow} diamonds, which correspond to experimentally measured lattice constants of 12~nm films (i.e., partially strained). The magnetic properties in these phase diagrams are calculated entirely through \textit{ab initio} methods, i.e., our experimental data refers to the lattice constants only.}  
We note that these calculations were performed without including the Hubbard $U$ correction, and the possible influence of this parameter will be discussed later.  To quantitatively describe its altermagnetic properties, here, we defined the N\'eel vector magnitude as $N = |\vec{\mu}_{\rm{Ru_1}}-\vec{\mu}_{\rm{Ru_2}}|/2$ where $\vec{\mu}_{\rm{Ru_1}}$ and $\vec{\mu}_{\rm{Ru_2}}$ represent the two opposite local magnetic moments of the Ru atoms in the primitive cell.
\textcolor{black}{Additional details of the first-principles calculations are provided in the SI.~\cite{SI}}


For (001) RuO$_2$, we find that sufficiently large strains along the [100] and [010] axes can stabilize the altermagnetic phase, as shown in Fig.~\ref{Fig1}(a). However, the fully strained (001) RuO$_2$/TiO$_2$ film \textcolor{black}{lattice constants} lie within the nonmagnetic region of the phase diagram, indicating the absence of magnetism in RuO$_2$ grown on (001) TiO$_2$. By contrast, as shown in Figs.~\ref{Fig1}(b) and (c), strong compressive strain along the [001] direction predominantly stabilizes finite magnetic moments in RuO$_2$, which can be realized in fully strained (100) and (110) films on TiO$_2$. These results are consistent with experimental observations of altermagnetic order in RuO$_2$ thin films,~\cite{jeong2024altermagnetic,jeong2025metallicity} even without including any Hubbard $U$ correction. Moreover, tensile strains along [100] and [1$\bar{1}$0] directions, respectively, further enhance the magnetic moments. This secondary effect can be attributed to piezomagnetism in the rutile crystal structure.~\cite{moriya1959piezomagnetism,komuro2025revisiting} However, the N\'eel vector vanishes in 12~nm RuO$_2$ films due to strain relaxation along [001]. These results underscore the fragile nature of magnetism in RuO$_2$ thin films,~\cite{SI} offering important design rules in stabilizing the magnetic ground state. \textcolor{black}{Our XRD results indicate that, despite thickness-dependent changes in peak intensity and broadening, the lattice constants of strain-relaxed RuO$_2$ layers grown on (100) TiO$_2$ exhibit small variations between 7 and 13 nm,~\cite{SI} consistent with previous reports using e-beam-assisted MBE.~\cite{https://doi.org/10.1038/s43246-025-00856-6}}

Beyond inducing local magnetism, epitaxial strain also alters the symmetry of the material. In particular, altermagnetism is characterized by \textcolor{black}{proper or improper} rotation, combined with time reversal that restores the original spin configuration. This symmetry is expressed by the spin–symmetry operation $\left [R_s||R_l \right ]$, where the left (right) operator acts in spin (real) space.~\cite{PhysRevX.12.040501,vsmejkal2022beyond,bai2024altermagnetism}
For RuO$_2$, both $\left [C_2||C_{4z} \tau \right ]$ and $\left [C_2||C_{2x} \tau \right ]$ symmetries satisfy the altermagnetic condition, where $C_2$ denotes spin-space inversion, $C_{4z}$ and $C_{2x}$ are spatial rotations, and $\tau$ represents a half-translation.
All these symmetries are preserved in fully strained (001) RuO$_2$, \textcolor{black}{which remains in space group $P4_2/mnm$}, as shown in Fig.~\ref{fig:Fig0}(a). However, the absence of local magnetic moments in this geometry leads to a nonmagnetic ground state. In contrast, for (100) RuO$_2$, the $\left [C_2||C_{4z}\tau \right ]$ symmetry is broken due to its orthorhombic lattice distortion. \textcolor{black}{In this case, the space group is the orthorhombic $Pnnm$, which results in invariance under }  $\left [C_2||C_{2x}\tau \right ]$. Together with the finite local magnetic moments arising from compressive strain, (100) RuO$_2$ yields the ideal altermagnetic phase, as shown in Fig.~\ref{fig:Fig0}(b). 
In fully strained (110) RuO$_2$, both altermagnetic symmetries are completely broken, giving rise to a finite net magnetic moment.~\cite{jeong2025metallicity} To examine this, we explore the strain–net magnetic moment phase diagram of (110) RuO$_2$, defined as $m=|\vec{\mu}_{Ru_1}+\vec{\mu}_{Ru_2}|$, in Fig.~\ref{Fig1}(d). \textcolor{black}{The net magnetization here is $\approx 0.025~\mu_B$. However, it} becomes larger in regions with high anisotropy between $a_{110}$ and $a_{1\bar{1}0}$. Thus, (110) RuO$_2$ is classified into a \textcolor{black}{uncompensated ferrimagnet}, which follows the same spin-space group configuration as a ferromagnet. 

\begin{figure}[t]
    \centering
    \includegraphics[width=\linewidth]{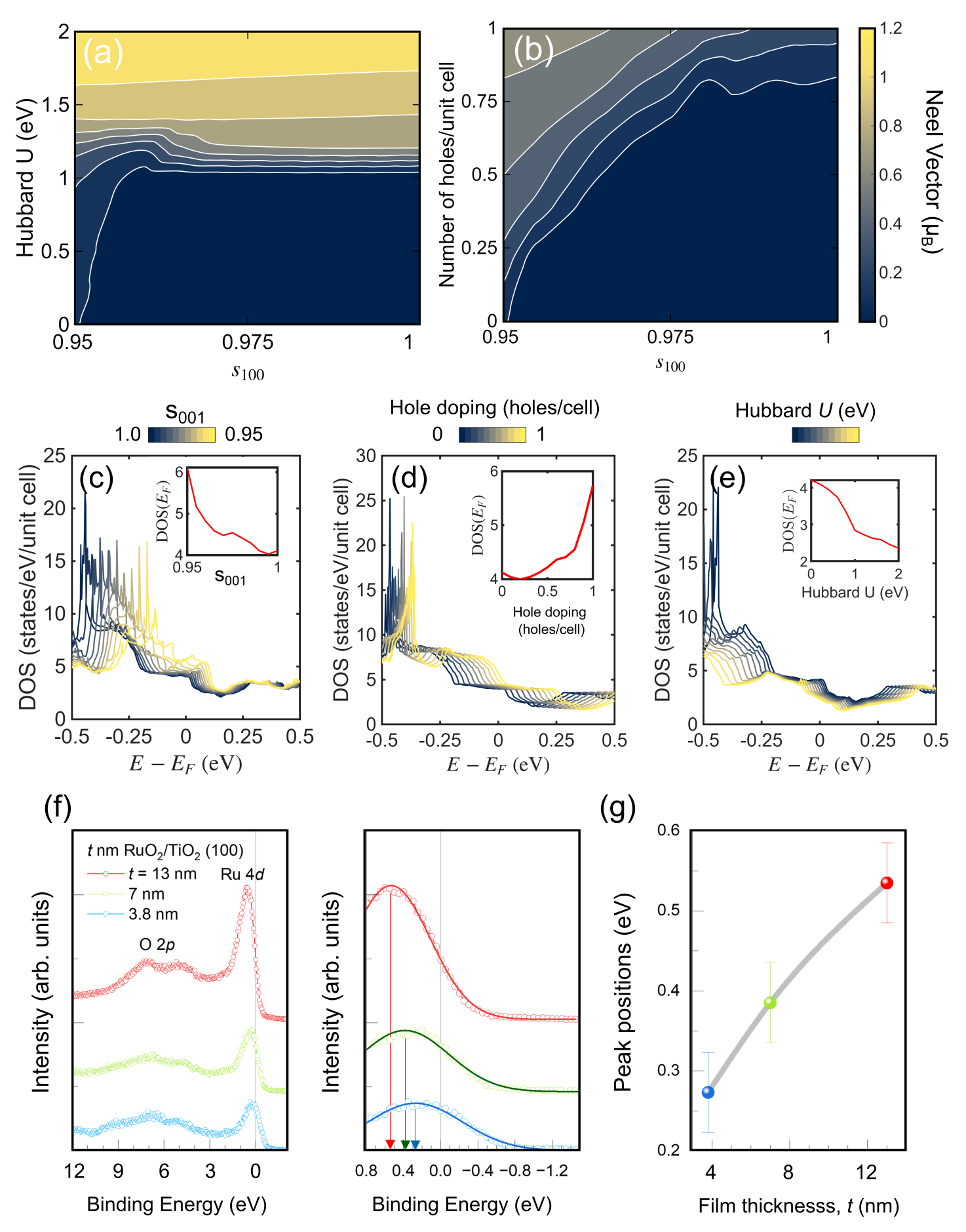}
    \caption{(a) Hubbard $U$ and (b) hole doping-strain phase diagrams for (100) RuO$_2$ showing the Neel vector. Density of states for (c) charge neutral (100) RuO$_2$ from pristine to fully strained, (d) unstrained (100) RuO$_2$ as a function of hole-doping and (e) unstrained, charge neutral (100) RuO$_2$ as a function of the Hubbard $U$ strength. The insets show the increase of the density of states at the Fermi level, which dictates the itinerant electron magnetism in (c) and (d), while the decreasing DOS($E_F$) in (e) reveals the localized character of this type of magnetism. (f) XPS valence band spectra of RuO$_2$ films on TiO$_2$ (100) substrates with different thickness $t$. The vertical arrows in the right panel indicate the estimated peak positions of the Ru $4d$ states, obtained by fitting with Voigt functions. (g) Summary of the Ru $4d$ peak positions as a function of $t$.}
    \label{fig:Fig2}
\end{figure}

\paragraph*{Origin of magnetism in RuO$_2$} 

Experimentally, altermagnetic signatures have also been reported in (001) and (101) RuO$_2$ thin films, indicating that the underlying physics extends beyond simple strain effects. To clarify the microscopic origin of magnetism in RuO$_2$, we construct extended magnetic phase diagrams including hole doping and Hubbard $U$ for (100) RuO$_2$, as shown in Fig.~\ref{fig:Fig2}(a) and (b), respectively. We find that both hole doping~\cite{PhysRevB.109.134424} and finite $U$ stabilize local magnetic order even without strain, thereby reducing the critical strain for the nonmagnetic–altermagnetic transition. For a given strain state, both Hubbard $U$ and hole doping increase the local magnetic moment. Thus, both electron correlation and hole doping can contribute to promote magnetism at low strain in experiments.
\textcolor{black}{Consistently, we also observe hole-doping–induced magnetism in (110) RuO$_2$, further supporting the proposed mechanism.~\cite{SI}}

To further understand the physical origin of magnetism in RuO$_2$, we analyze the strain-, doping-, and $U$-dependent nonmagnetic density of states (DOS), shown in Fig.~\ref{fig:Fig2}(c-e). For pristine RuO$_2$ with $U=0$~eV, the DOS near the Fermi level ($E_{\rm{F}}$) is relatively low, but a strong peak exists around $E=E_{\rm{F}}-0.45$~eV. Either compressive strain along [001] or hole doping shifts this peak toward $E_{\rm{F}}$, thereby enhancing the DOS at $E_{\rm{F}}$ and triggering spontaneous magnetization via Stoner instability.~\cite{PhysRevB.109.134424,jeong2025metallicity} In contrast, a finite $U$ shifts the DOS in the opposite direction. This indicates that near $U{\sim}0$~eV, itinerant magnetism arises from Fermi surface instability associated with high DOS, which is strain-sensitive, while larger $U$ drives strain-independent localized magnetism. 

At this point, it is important to discuss the reasonable magnitude of the Hubbard $U$ for RuO$_2$. Recent experiments consistently suggest that $U>1.2$~eV is highly unlikely. 
For example, the experimentally reported $|\vec{\mu}_{\rm{Ru}}|$ of 0.05–0.15~$\mu_{\rm B}$ is far below than the $\sim$1.0~$\mu_{\rm B}$ predicted by $U>1.2$~eV.~\cite{berlijn2017itinerant,zhu2019anomalous}
The absence of magnetism in high-purity bulk RuO$_2$ also supports a small $U$.~\cite{hiraishi2024nonmagnetic,kessler2024absence} 
Also, optical spectroscopy~\cite{wenzel2025fermi,jeong2025anisotropic} and spin-resolved ARPES~\cite{liu2024absence,yichen2025} further support $U{\sim}0$. Therefore, the observed magnetism of RuO$_2$ in experiments is most likely understood as arising from strain- or doping-induced itinerant magnetism, with only a minor role for $U$-driven localization. Nonetheless, it is worth noting that a small but finite $U$ is not fully excluded yet.

\begin{figure*}[t]
    \centering
    \includegraphics[width=\linewidth]{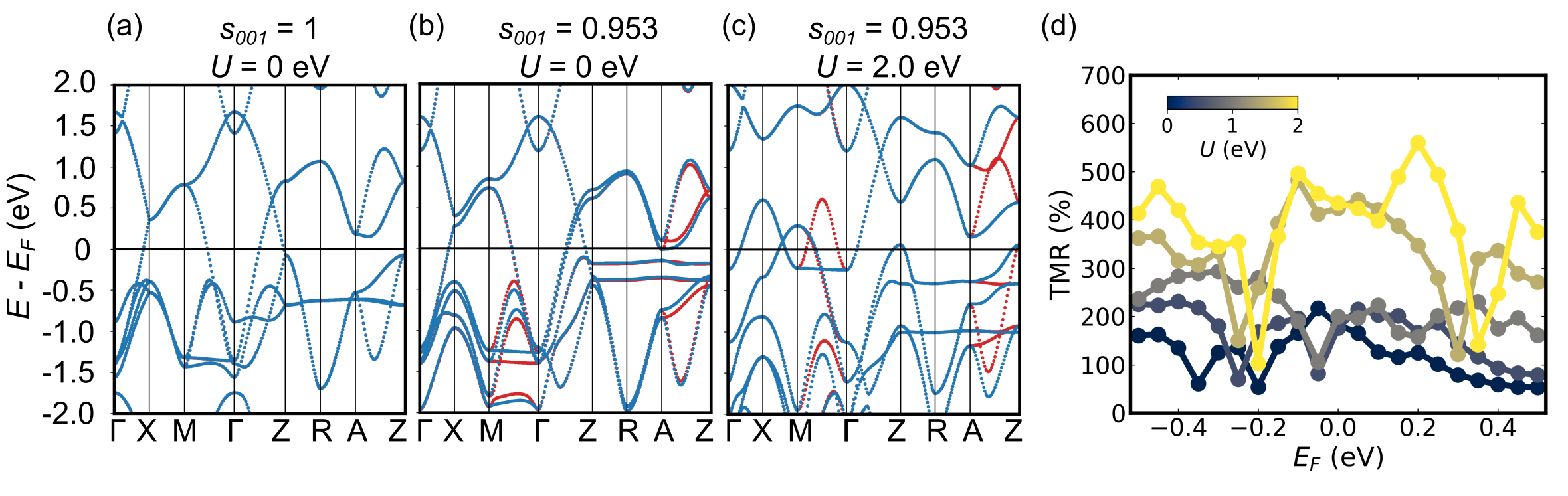}
    \caption{Band structures for (a) unstrained (100) RuO$_2$ with $U = 0$ eV and fully strained (100) RuO$_2$ with (b) $U = 0$ eV and (c) $U = 2$ eV. Strain alone can drive the spin-splitting of the bands, which is further enhanced by the Hubbard $U$ interaction. In (d), we evaluate the TMR as a function of the Fermi level for different $U$, showing that it increases for largeer spin-splittings, as expected.}
    \label{Fig3}
\end{figure*}

To validate the theoretically proposed strain scenario, we synthesized epitaxial RuO$_2$ films on TiO$_2$ (100) substrates with thicknesses of 3.8, 7, and 13~nm using hybrid molecular beam epitaxy (hMBE) (Figs.~S1a–c and S2).~\cite{SI} Structural characterizations by \textcolor{black}{x-ray reflectometry (}XRR\textcolor{black}{)} and XRD (Fig.~S2) confirm high crystalline quality. XRD reciprocal space maps (RSMs) around the TiO$_2$ (30$\bar{1}$) reflection reveal that the 3.8~nm film remains fully strained, while the 7~nm film shows a partially relaxed peak near $Q_{001} \approx -3.28~\text{\AA}^{-1}$, which becomes more pronounced for the 13~nm film.~\cite{SI} These results indicate a critical thickness for strain relaxation along [001] of $\sim$4~nm, consistent with previous reports on RuO$_2$/TiO$_2$ (110) heterostructures~\cite{jeong2025strainstabilizedinterfacialpolarizationtunes, jeong2025anisotropic, jeong2025metallicity}. The in-plane [010] lattice parameter remains coherently strained up to 13~nm (Table~S1), suggesting that the strain effect primarily arises along the [001] direction.

Strain-dependent electronic structure modulation near the Fermi level was further examined using valence-band x-ray photoelectron spectroscopy (XPS) (Figs.~\ref{fig:Fig2}(f)-(g), S4).~\cite{SI} The spectra display O~2$p$ states between 2~eV and 10~eV binding energy and Ru~4$d$ states from 0 to 2~eV (Fig.~\ref{fig:Fig2}(f), left)~\cite{Stoeberl_2020}. The right panel of Fig.~\ref{fig:Fig2}(f) highlights the evolution of the Ru~4$d$ feature near the Fermi level. The spectra were fitted with Voigt functions to extract peak positions, summarized in Fig.~\ref{fig:Fig2}(g). The Ru~4$d$ peak systematically shifts toward the Fermi level as the film thickness decreases from 13~nm to 3.8~nm. This observation demonstrates that epitaxial strain along [001] drives the Ru~4$d$ valence state closer to the Fermi level~\cite{jeong2025anisotropic}, in agreement with our theoretical calculations.

\paragraph*{Re-evaluation of electronic structure of \textcolor{black}{(100)} RuO$_2$} To re-evaluate the electronic structure of RuO$_2$, we compared the band structures of pristine, strained, and \textcolor{black}{combined }strained with Hubbard $U$ RuO$_2$, as shown in Fig.~\ref{Fig3}(a–c). Without $U$, the band structure of altermagnetic (strained) RuO$_2$ closely resembles that of the nonmagnetic (pristine) phase. Under compressive strain, the flat band along Z–R–A–Z shifts toward $E_{\rm{F}}$, thereby enhancing the density of states. The spin splitting energy in (100) RuO$_2$ is on the order of several tens of meV, which is much smaller than the 1.4~eV predicted for $U=2$~eV,~\cite{PhysRevX.12.040501} mainly along the M$-\Gamma$ line.

Finally, we evaluate tunneling magnetoresistance (TMR) of magnetic tunnel junction (MTJ) based on (100) RuO$_2$. To estimate the TMR, we developed a simple model based on Julliere's approach,~\cite{julliere1975tunneling} defined as ${\rm{TMR}} \equiv ({R_{ap} - R_p})/{R_p}$,
where $R_p$ and $R_{ap}$ are the resistances in parallel and anti-parallel state configurations, respectively. 
The resistances are in turn estimated through the overlap of the Fermi surfaces, calculated using the spin-dependent spectral function $A_\sigma (\mathbf{k}, \omega) = \frac{1}{\pi} \text{Im}\left(\hbar \omega - \epsilon_{\mathbf{k}, \sigma} - i \eta\right)^{-1}$ as follows

\begin{align}
    K_{\sigma \sigma'}(\omega)&=\frac{\int_{BZ} d^3k A_\sigma(\mathbf{k},\omega) A_{\sigma'} (\mathbf{k},\omega)}{\int_{BZ} d^3 k \left[ \abs{A_{\uparrow} (\mathbf{k},\omega)}^2 + \abs{A_\downarrow (\mathbf{k},\omega)}^2\right]},
\end{align}
where $\sigma = \uparrow,\downarrow$ is the spin index and the broadening factor of $\eta$ is chosen to be 1~meV.
Then, we simply assume that the resistances are proportional to the inverse of the overlap $K_{\sigma \sigma'}(\omega)$, $R_{p}= 1/{K_{\uparrow \uparrow}}$ and $R_{ap} = 1/{K_{\uparrow \downarrow}}$.
We emphasize that this model ignores the microscopic structure of the insulating barrier.~\cite{julliere1975tunneling}

Our results are summarized in Fig.~\ref{Fig3}(d). For $U<$~1.2~eV, the TMR at $E_{\rm{F}}$ is nearly insensitive to $U$ and remains \textcolor{black}{at a maximum of approximately 200~\%}. With larger $U$, strong spin splitting along M–$\Gamma$ at $E_{\rm{F}}$ emerges, enhancing the TMR up to $\sim$\textcolor{black}{600}~\%. However, as discussed earlier, such strong $U$ values are \textcolor{black}{likely unrealistic} and may lead to an overestimation of TMR. Nevertheless, the TMR can be further improved by selecting an appropriate insulating barrier that minimizes $R_p$ through optimizing momentum-dependent spin-polarized conduction.~\cite{jiang2023prediction,gurung2024nearly}
The detailed spin-resolved Fermi surface of (100) RuO$_2$ is provided in Supplementary Information.~\cite{SI} We further note that a nonzero TMR is also expected in (110) RuO$_2$, but not in (001) RuO$_2$ due to the absence of local magnetism.

\paragraph*{Conclusion} In summary, combining XRD measurements with first-principles calculations, we demonstrate that epitaxial strain plays a crucial role in RuO$_2$ thin films by manipulating their magnetic ground state. Compressive strain along [001] enhances the density of states near $E{_{\rm{F}}}$, driving Fermi surface instabilities and itinerant altermagnetism. Experimentally, (100) and (110) RuO$_2$ films grown on TiO$_2$ substrates satisfy these conditions, with the former exhibiting ideal altermagnetism and the latter showing ferrimagnetism due to broken rotation symmetries. We re-evalute its spin-splitting energy and corresponding TMR of a (100) RuO$_2$-based MTJ structure, both of which are found to be smaller than previously reported values. Our results provide a fundamental understanding of the origin of altermagnetism in RuO$_2$ and thus reconcile the ongoing debate surrounding its magnetic nature.

\paragraph*{Acknowledgements} This work was supported by AFOSR sponsored MURI (Grant \#FA9550-25-1-0262).


\begin{thebibliography}{47}%
\makeatletter
\providecommand \@ifxundefined [1]{%
 \@ifx{#1\undefined}
}%
\providecommand \@ifnum [1]{%
 \ifnum #1\expandafter \@firstoftwo
 \else \expandafter \@secondoftwo
 \fi
}%
\providecommand \@ifx [1]{%
 \ifx #1\expandafter \@firstoftwo
 \else \expandafter \@secondoftwo
 \fi
}%
\providecommand \natexlab [1]{#1}%
\providecommand \enquote  [1]{``#1''}%
\providecommand \bibnamefont  [1]{#1}%
\providecommand \bibfnamefont [1]{#1}%
\providecommand \citenamefont [1]{#1}%
\providecommand \href@noop [0]{\@secondoftwo}%
\providecommand \href [0]{\begingroup \@sanitize@url \@href}%
\providecommand \@href[1]{\@@startlink{#1}\@@href}%
\providecommand \@@href[1]{\endgroup#1\@@endlink}%
\providecommand \@sanitize@url [0]{\catcode `\\12\catcode `\$12\catcode `\&12\catcode `\#12\catcode `\^12\catcode `\_12\catcode `\%12\relax}%
\providecommand \@@startlink[1]{}%
\providecommand \@@endlink[0]{}%
\providecommand \url  [0]{\begingroup\@sanitize@url \@url }%
\providecommand \@url [1]{\endgroup\@href {#1}{\urlprefix }}%
\providecommand \urlprefix  [0]{URL }%
\providecommand \Eprint [0]{\href }%
\providecommand \doibase [0]{http://dx.doi.org/}%
\providecommand \selectlanguage [0]{\@gobble}%
\providecommand \bibinfo  [0]{\@secondoftwo}%
\providecommand \bibfield  [0]{\@secondoftwo}%
\providecommand \translation [1]{[#1]}%
\providecommand \BibitemOpen [0]{}%
\providecommand \bibitemStop [0]{}%
\providecommand \bibitemNoStop [0]{.\EOS\space}%
\providecommand \EOS [0]{\spacefactor3000\relax}%
\providecommand \BibitemShut  [1]{\csname bibitem#1\endcsname}%
\let\auto@bib@innerbib\@empty
\bibitem [{\citenamefont {Šmejkal}\ \emph {et~al.}(2020)\citenamefont {Šmejkal}, \citenamefont {González-Hernández}, \citenamefont {Jungwirth},\ and\ \citenamefont {Sinova}}]{10.1126/sciadv.aaz8809}%
  \BibitemOpen
  \bibfield  {author} {\bibinfo {author} {\bibfnamefont {L.}~\bibnamefont {Šmejkal}}, \bibinfo {author} {\bibfnamefont {R.}~\bibnamefont {González-Hernández}}, \bibinfo {author} {\bibfnamefont {T.}~\bibnamefont {Jungwirth}}, \ and\ \bibinfo {author} {\bibfnamefont {J.}~\bibnamefont {Sinova}},\ }\href {\doibase 10.1126/sciadv.aaz8809} {\bibfield  {journal} {\bibinfo  {journal} {Sci. Adv.}\ }\textbf {\bibinfo {volume} {6}},\ \bibinfo {pages} {eaaz8809} (\bibinfo {year} {2020})}\BibitemShut {NoStop}%
\bibitem [{\citenamefont {\ifmmode~\check{S}\else \v{S}\fi{}mejkal}\ \emph {et~al.}(2022)\citenamefont {\ifmmode~\check{S}\else \v{S}\fi{}mejkal}, \citenamefont {Sinova},\ and\ \citenamefont {Jungwirth}}]{PhysRevX.12.040501}%
  \BibitemOpen
  \bibfield  {author} {\bibinfo {author} {\bibfnamefont {L.}~\bibnamefont {\ifmmode~\check{S}\else \v{S}\fi{}mejkal}}, \bibinfo {author} {\bibfnamefont {J.}~\bibnamefont {Sinova}}, \ and\ \bibinfo {author} {\bibfnamefont {T.}~\bibnamefont {Jungwirth}},\ }\href {\doibase 10.1103/PhysRevX.12.040501} {\bibfield  {journal} {\bibinfo  {journal} {Phys. Rev. X}\ }\textbf {\bibinfo {volume} {12}},\ \bibinfo {pages} {040501} (\bibinfo {year} {2022})}\BibitemShut {NoStop}%
\bibitem [{\citenamefont {{\v{S}}mejkal}\ \emph {et~al.}(2022)\citenamefont {{\v{S}}mejkal}, \citenamefont {Sinova},\ and\ \citenamefont {Jungwirth}}]{vsmejkal2022beyond}%
  \BibitemOpen
  \bibfield  {author} {\bibinfo {author} {\bibfnamefont {L.}~\bibnamefont {{\v{S}}mejkal}}, \bibinfo {author} {\bibfnamefont {J.}~\bibnamefont {Sinova}}, \ and\ \bibinfo {author} {\bibfnamefont {T.}~\bibnamefont {Jungwirth}},\ }\href@noop {} {\bibfield  {journal} {\bibinfo  {journal} {Phys. Rev. X}\ }\textbf {\bibinfo {volume} {12}},\ \bibinfo {pages} {031042} (\bibinfo {year} {2022})}\BibitemShut {NoStop}%
\bibitem [{\citenamefont {Yuan}\ \emph {et~al.}(2021)\citenamefont {Yuan}, \citenamefont {Wang}, \citenamefont {Luo},\ and\ \citenamefont {Zunger}}]{yuan2021prediction}%
  \BibitemOpen
  \bibfield  {author} {\bibinfo {author} {\bibfnamefont {L.-D.}\ \bibnamefont {Yuan}}, \bibinfo {author} {\bibfnamefont {Z.}~\bibnamefont {Wang}}, \bibinfo {author} {\bibfnamefont {J.-W.}\ \bibnamefont {Luo}}, \ and\ \bibinfo {author} {\bibfnamefont {A.}~\bibnamefont {Zunger}},\ }\href@noop {} {\bibfield  {journal} {\bibinfo  {journal} {Phys. Rev. Mater.}\ }\textbf {\bibinfo {volume} {5}},\ \bibinfo {pages} {014409} (\bibinfo {year} {2021})}\BibitemShut {NoStop}%
\bibitem [{\citenamefont {Shao}\ \emph {et~al.}(2021)\citenamefont {Shao}, \citenamefont {Zhang}, \citenamefont {Li}, \citenamefont {Eom},\ and\ \citenamefont {Tsymbal}}]{shao2021spin}%
  \BibitemOpen
  \bibfield  {author} {\bibinfo {author} {\bibfnamefont {D.-F.}\ \bibnamefont {Shao}}, \bibinfo {author} {\bibfnamefont {S.-H.}\ \bibnamefont {Zhang}}, \bibinfo {author} {\bibfnamefont {M.}~\bibnamefont {Li}}, \bibinfo {author} {\bibfnamefont {C.-B.}\ \bibnamefont {Eom}}, \ and\ \bibinfo {author} {\bibfnamefont {E.~Y.}\ \bibnamefont {Tsymbal}},\ }\href@noop {} {\bibfield  {journal} {\bibinfo  {journal} {Nat. Commun.}\ }\textbf {\bibinfo {volume} {12}},\ \bibinfo {pages} {7061} (\bibinfo {year} {2021})}\BibitemShut {NoStop}%
\bibitem [{\citenamefont {{\v{S}}mejkal}\ \emph {et~al.}(2023)\citenamefont {{\v{S}}mejkal}, \citenamefont {Marmodoro}, \citenamefont {Ahn}, \citenamefont {Gonz{\'a}lez-Hern{\'a}ndez}, \citenamefont {Turek}, \citenamefont {Mankovsky}, \citenamefont {Ebert}, \citenamefont {D’Souza}, \citenamefont {{\v{S}}ipr}, \citenamefont {Sinova} \emph {et~al.}}]{vsmejkal2023chiral}%
  \BibitemOpen
  \bibfield  {author} {\bibinfo {author} {\bibfnamefont {L.}~\bibnamefont {{\v{S}}mejkal}}, \bibinfo {author} {\bibfnamefont {A.}~\bibnamefont {Marmodoro}}, \bibinfo {author} {\bibfnamefont {K.-H.}\ \bibnamefont {Ahn}}, \bibinfo {author} {\bibfnamefont {R.}~\bibnamefont {Gonz{\'a}lez-Hern{\'a}ndez}}, \bibinfo {author} {\bibfnamefont {I.}~\bibnamefont {Turek}}, \bibinfo {author} {\bibfnamefont {S.}~\bibnamefont {Mankovsky}}, \bibinfo {author} {\bibfnamefont {H.}~\bibnamefont {Ebert}}, \bibinfo {author} {\bibfnamefont {S.~W.}\ \bibnamefont {D’Souza}}, \bibinfo {author} {\bibfnamefont {O.}~\bibnamefont {{\v{S}}ipr}}, \bibinfo {author} {\bibfnamefont {J.}~\bibnamefont {Sinova}},  \emph {et~al.},\ }\href@noop {} {\bibfield  {journal} {\bibinfo  {journal} {Phys. Rev. Lett.}\ }\textbf {\bibinfo {volume} {131}},\ \bibinfo {pages} {256703} (\bibinfo {year} {2023})}\BibitemShut {NoStop}%
\bibitem [{\citenamefont {Bai}\ \emph {et~al.}(2024)\citenamefont {Bai}, \citenamefont {Feng}, \citenamefont {Liu}, \citenamefont {{\v{S}}mejkal}, \citenamefont {Mokrousov},\ and\ \citenamefont {Yao}}]{bai2024altermagnetism}%
  \BibitemOpen
  \bibfield  {author} {\bibinfo {author} {\bibfnamefont {L.}~\bibnamefont {Bai}}, \bibinfo {author} {\bibfnamefont {W.}~\bibnamefont {Feng}}, \bibinfo {author} {\bibfnamefont {S.}~\bibnamefont {Liu}}, \bibinfo {author} {\bibfnamefont {L.}~\bibnamefont {{\v{S}}mejkal}}, \bibinfo {author} {\bibfnamefont {Y.}~\bibnamefont {Mokrousov}}, \ and\ \bibinfo {author} {\bibfnamefont {Y.}~\bibnamefont {Yao}},\ }\href@noop {} {\bibfield  {journal} {\bibinfo  {journal} {Adv. Funct. Mater.}\ }\textbf {\bibinfo {volume} {34}},\ \bibinfo {pages} {2409327} (\bibinfo {year} {2024})}\BibitemShut {NoStop}%
\bibitem [{\citenamefont {Berlijn}\ \emph {et~al.}(2017{\natexlab{a}})\citenamefont {Berlijn}, \citenamefont {Snijders}, \citenamefont {Delaire}, \citenamefont {Zhou}, \citenamefont {Maier}, \citenamefont {Cao}, \citenamefont {Chi}, \citenamefont {Matsuda}, \citenamefont {Wang}, \citenamefont {Koehler} \emph {et~al.}}]{berlijn2017itinerant}%
  \BibitemOpen
  \bibfield  {author} {\bibinfo {author} {\bibfnamefont {T.}~\bibnamefont {Berlijn}}, \bibinfo {author} {\bibfnamefont {P.~C.}\ \bibnamefont {Snijders}}, \bibinfo {author} {\bibfnamefont {O.}~\bibnamefont {Delaire}}, \bibinfo {author} {\bibfnamefont {H.-D.}\ \bibnamefont {Zhou}}, \bibinfo {author} {\bibfnamefont {T.~A.}\ \bibnamefont {Maier}}, \bibinfo {author} {\bibfnamefont {H.-B.}\ \bibnamefont {Cao}}, \bibinfo {author} {\bibfnamefont {S.-X.}\ \bibnamefont {Chi}}, \bibinfo {author} {\bibfnamefont {M.}~\bibnamefont {Matsuda}}, \bibinfo {author} {\bibfnamefont {Y.}~\bibnamefont {Wang}}, \bibinfo {author} {\bibfnamefont {M.~R.}\ \bibnamefont {Koehler}},  \emph {et~al.},\ }\href@noop {} {\bibfield  {journal} {\bibinfo  {journal} {Phys. Rev. Lett.}\ }\textbf {\bibinfo {volume} {118}},\ \bibinfo {pages} {077201} (\bibinfo {year} {2017}{\natexlab{a}})}\BibitemShut {NoStop}%
\bibitem [{\citenamefont {Krempask{\`y}}\ \emph {et~al.}(2024)\citenamefont {Krempask{\`y}}, \citenamefont {{\v{S}}mejkal}, \citenamefont {D’souza}, \citenamefont {Hajlaoui}, \citenamefont {Springholz}, \citenamefont {Uhl{\'\i}{\v{r}}ov{\'a}}, \citenamefont {Alarab}, \citenamefont {Constantinou}, \citenamefont {Strocov}, \citenamefont {Usanov} \emph {et~al.}}]{krempasky2024altermagnetic}%
  \BibitemOpen
  \bibfield  {author} {\bibinfo {author} {\bibfnamefont {J.}~\bibnamefont {Krempask{\`y}}}, \bibinfo {author} {\bibfnamefont {L.}~\bibnamefont {{\v{S}}mejkal}}, \bibinfo {author} {\bibfnamefont {S.}~\bibnamefont {D’souza}}, \bibinfo {author} {\bibfnamefont {M.}~\bibnamefont {Hajlaoui}}, \bibinfo {author} {\bibfnamefont {G.}~\bibnamefont {Springholz}}, \bibinfo {author} {\bibfnamefont {K.}~\bibnamefont {Uhl{\'\i}{\v{r}}ov{\'a}}}, \bibinfo {author} {\bibfnamefont {F.}~\bibnamefont {Alarab}}, \bibinfo {author} {\bibfnamefont {P.}~\bibnamefont {Constantinou}}, \bibinfo {author} {\bibfnamefont {V.}~\bibnamefont {Strocov}}, \bibinfo {author} {\bibfnamefont {D.}~\bibnamefont {Usanov}},  \emph {et~al.},\ }\href@noop {} {\bibfield  {journal} {\bibinfo  {journal} {Nature}\ }\textbf {\bibinfo {volume} {626}},\ \bibinfo {pages} {517} (\bibinfo {year} {2024})}\BibitemShut {NoStop}%
\bibitem [{\citenamefont {Lee}\ \emph {et~al.}(2024)\citenamefont {Lee}, \citenamefont {Lee}, \citenamefont {Jung}, \citenamefont {Jung}, \citenamefont {Kim}, \citenamefont {Lee}, \citenamefont {Seok}, \citenamefont {Kim}, \citenamefont {Park}, \citenamefont {{\v{S}}mejkal} \emph {et~al.}}]{lee2024broken}%
  \BibitemOpen
  \bibfield  {author} {\bibinfo {author} {\bibfnamefont {S.}~\bibnamefont {Lee}}, \bibinfo {author} {\bibfnamefont {S.}~\bibnamefont {Lee}}, \bibinfo {author} {\bibfnamefont {S.}~\bibnamefont {Jung}}, \bibinfo {author} {\bibfnamefont {J.}~\bibnamefont {Jung}}, \bibinfo {author} {\bibfnamefont {D.}~\bibnamefont {Kim}}, \bibinfo {author} {\bibfnamefont {Y.}~\bibnamefont {Lee}}, \bibinfo {author} {\bibfnamefont {B.}~\bibnamefont {Seok}}, \bibinfo {author} {\bibfnamefont {J.}~\bibnamefont {Kim}}, \bibinfo {author} {\bibfnamefont {B.~G.}\ \bibnamefont {Park}}, \bibinfo {author} {\bibfnamefont {L.}~\bibnamefont {{\v{S}}mejkal}},  \emph {et~al.},\ }\href@noop {} {\bibfield  {journal} {\bibinfo  {journal} {Phys. Rev. Lett.}\ }\textbf {\bibinfo {volume} {132}},\ \bibinfo {pages} {036702} (\bibinfo {year} {2024})}\BibitemShut {NoStop}%
\bibitem [{\citenamefont {Zhou}\ \emph {et~al.}(2025)\citenamefont {Zhou}, \citenamefont {Cheng}, \citenamefont {Hu}, \citenamefont {Chu}, \citenamefont {Bai}, \citenamefont {Han}, \citenamefont {Liu}, \citenamefont {Pan},\ and\ \citenamefont {Song}}]{zhou2025manipulation}%
  \BibitemOpen
  \bibfield  {author} {\bibinfo {author} {\bibfnamefont {Z.}~\bibnamefont {Zhou}}, \bibinfo {author} {\bibfnamefont {X.}~\bibnamefont {Cheng}}, \bibinfo {author} {\bibfnamefont {M.}~\bibnamefont {Hu}}, \bibinfo {author} {\bibfnamefont {R.}~\bibnamefont {Chu}}, \bibinfo {author} {\bibfnamefont {H.}~\bibnamefont {Bai}}, \bibinfo {author} {\bibfnamefont {L.}~\bibnamefont {Han}}, \bibinfo {author} {\bibfnamefont {J.}~\bibnamefont {Liu}}, \bibinfo {author} {\bibfnamefont {F.}~\bibnamefont {Pan}}, \ and\ \bibinfo {author} {\bibfnamefont {C.}~\bibnamefont {Song}},\ }\href@noop {} {\bibfield  {journal} {\bibinfo  {journal} {Nature}\ ,\ \bibinfo {pages} {1}} (\bibinfo {year} {2025})}\BibitemShut {NoStop}%
\bibitem [{\citenamefont {Han}\ \emph {et~al.}(2024)\citenamefont {Han}, \citenamefont {Fu}, \citenamefont {Peng}, \citenamefont {Cheng}, \citenamefont {Dai}, \citenamefont {Liu}, \citenamefont {Li}, \citenamefont {Zhang}, \citenamefont {Zhu}, \citenamefont {Bai} \emph {et~al.}}]{han2024electrical}%
  \BibitemOpen
  \bibfield  {author} {\bibinfo {author} {\bibfnamefont {L.}~\bibnamefont {Han}}, \bibinfo {author} {\bibfnamefont {X.}~\bibnamefont {Fu}}, \bibinfo {author} {\bibfnamefont {R.}~\bibnamefont {Peng}}, \bibinfo {author} {\bibfnamefont {X.}~\bibnamefont {Cheng}}, \bibinfo {author} {\bibfnamefont {J.}~\bibnamefont {Dai}}, \bibinfo {author} {\bibfnamefont {L.}~\bibnamefont {Liu}}, \bibinfo {author} {\bibfnamefont {Y.}~\bibnamefont {Li}}, \bibinfo {author} {\bibfnamefont {Y.}~\bibnamefont {Zhang}}, \bibinfo {author} {\bibfnamefont {W.}~\bibnamefont {Zhu}}, \bibinfo {author} {\bibfnamefont {H.}~\bibnamefont {Bai}},  \emph {et~al.},\ }\href@noop {} {\bibfield  {journal} {\bibinfo  {journal} {Sci. Adv.}\ }\textbf {\bibinfo {volume} {10}},\ \bibinfo {pages} {eadn0479} (\bibinfo {year} {2024})}\BibitemShut {NoStop}%
\bibitem [{\citenamefont {Takagi}\ \emph {et~al.}(2025)\citenamefont {Takagi}, \citenamefont {Hirakida}, \citenamefont {Settai}, \citenamefont {Oiwa}, \citenamefont {Takagi}, \citenamefont {Kitaori}, \citenamefont {Yamauchi}, \citenamefont {Inoue}, \citenamefont {Yamaura}, \citenamefont {Nishio-Hamane} \emph {et~al.}}]{takagi2025spontaneous}%
  \BibitemOpen
  \bibfield  {author} {\bibinfo {author} {\bibfnamefont {R.}~\bibnamefont {Takagi}}, \bibinfo {author} {\bibfnamefont {R.}~\bibnamefont {Hirakida}}, \bibinfo {author} {\bibfnamefont {Y.}~\bibnamefont {Settai}}, \bibinfo {author} {\bibfnamefont {R.}~\bibnamefont {Oiwa}}, \bibinfo {author} {\bibfnamefont {H.}~\bibnamefont {Takagi}}, \bibinfo {author} {\bibfnamefont {A.}~\bibnamefont {Kitaori}}, \bibinfo {author} {\bibfnamefont {K.}~\bibnamefont {Yamauchi}}, \bibinfo {author} {\bibfnamefont {H.}~\bibnamefont {Inoue}}, \bibinfo {author} {\bibfnamefont {J.-i.}\ \bibnamefont {Yamaura}}, \bibinfo {author} {\bibfnamefont {D.}~\bibnamefont {Nishio-Hamane}},  \emph {et~al.},\ }\href@noop {} {\bibfield  {journal} {\bibinfo  {journal} {Nat. Mater.}\ }\textbf {\bibinfo {volume} {24}},\ \bibinfo {pages} {63} (\bibinfo {year} {2025})}\BibitemShut {NoStop}%
\bibitem [{\citenamefont {Jiang}\ \emph {et~al.}(2025)\citenamefont {Jiang}, \citenamefont {Hu}, \citenamefont {Bai}, \citenamefont {Song}, \citenamefont {Mu}, \citenamefont {Qu}, \citenamefont {Li}, \citenamefont {Zhu}, \citenamefont {Pi}, \citenamefont {Wei}, \citenamefont {Sun}, \citenamefont {Huang}, \citenamefont {Zheng}, \citenamefont {Peng}, \citenamefont {He}, \citenamefont {Li}, \citenamefont {Luo}, \citenamefont {Li}, \citenamefont {Chen}, \citenamefont {Li}, \citenamefont {Weng},\ and\ \citenamefont {Qian}}]{Jiang2025}%
  \BibitemOpen
  \bibfield  {author} {\bibinfo {author} {\bibfnamefont {B.}~\bibnamefont {Jiang}}, \bibinfo {author} {\bibfnamefont {M.}~\bibnamefont {Hu}}, \bibinfo {author} {\bibfnamefont {J.}~\bibnamefont {Bai}}, \bibinfo {author} {\bibfnamefont {Z.}~\bibnamefont {Song}}, \bibinfo {author} {\bibfnamefont {C.}~\bibnamefont {Mu}}, \bibinfo {author} {\bibfnamefont {G.}~\bibnamefont {Qu}}, \bibinfo {author} {\bibfnamefont {W.}~\bibnamefont {Li}}, \bibinfo {author} {\bibfnamefont {W.}~\bibnamefont {Zhu}}, \bibinfo {author} {\bibfnamefont {H.}~\bibnamefont {Pi}}, \bibinfo {author} {\bibfnamefont {Z.}~\bibnamefont {Wei}}, \bibinfo {author} {\bibfnamefont {Y.-J.}\ \bibnamefont {Sun}}, \bibinfo {author} {\bibfnamefont {Y.}~\bibnamefont {Huang}}, \bibinfo {author} {\bibfnamefont {X.}~\bibnamefont {Zheng}}, \bibinfo {author} {\bibfnamefont {Y.}~\bibnamefont {Peng}}, \bibinfo {author} {\bibfnamefont {L.}~\bibnamefont {He}}, \bibinfo {author} {\bibfnamefont {S.}~\bibnamefont {Li}}, \bibinfo {author} {\bibfnamefont {J.}~\bibnamefont
  {Luo}}, \bibinfo {author} {\bibfnamefont {Z.}~\bibnamefont {Li}}, \bibinfo {author} {\bibfnamefont {G.}~\bibnamefont {Chen}}, \bibinfo {author} {\bibfnamefont {H.}~\bibnamefont {Li}}, \bibinfo {author} {\bibfnamefont {H.}~\bibnamefont {Weng}}, \ and\ \bibinfo {author} {\bibfnamefont {T.}~\bibnamefont {Qian}},\ }\href {\doibase 10.1038/s41567-025-02822-y} {\bibfield  {journal} {\bibinfo  {journal} {Nat. Phys.}\ }\textbf {\bibinfo {volume} {21}},\ \bibinfo {pages} {754} (\bibinfo {year} {2025})}\BibitemShut {NoStop}%
\bibitem [{\citenamefont {Zhang}\ \emph {et~al.}(2025{\natexlab{a}})\citenamefont {Zhang}, \citenamefont {Cheng}, \citenamefont {Yin}, \citenamefont {Liu}, \citenamefont {Deng}, \citenamefont {Qiao}, \citenamefont {Shi}, \citenamefont {Zhang}, \citenamefont {Lin}, \citenamefont {Liu}, \citenamefont {Ye}, \citenamefont {Huang}, \citenamefont {Meng}, \citenamefont {Zhang}, \citenamefont {Okuda}, \citenamefont {Shimada}, \citenamefont {Cui}, \citenamefont {Zhao}, \citenamefont {Cao}, \citenamefont {Qiao}, \citenamefont {Liu},\ and\ \citenamefont {Chen}}]{Zhang2025}%
  \BibitemOpen
  \bibfield  {author} {\bibinfo {author} {\bibfnamefont {F.}~\bibnamefont {Zhang}}, \bibinfo {author} {\bibfnamefont {X.}~\bibnamefont {Cheng}}, \bibinfo {author} {\bibfnamefont {Z.}~\bibnamefont {Yin}}, \bibinfo {author} {\bibfnamefont {C.}~\bibnamefont {Liu}}, \bibinfo {author} {\bibfnamefont {L.}~\bibnamefont {Deng}}, \bibinfo {author} {\bibfnamefont {Y.}~\bibnamefont {Qiao}}, \bibinfo {author} {\bibfnamefont {Z.}~\bibnamefont {Shi}}, \bibinfo {author} {\bibfnamefont {S.}~\bibnamefont {Zhang}}, \bibinfo {author} {\bibfnamefont {J.}~\bibnamefont {Lin}}, \bibinfo {author} {\bibfnamefont {Z.}~\bibnamefont {Liu}}, \bibinfo {author} {\bibfnamefont {M.}~\bibnamefont {Ye}}, \bibinfo {author} {\bibfnamefont {Y.}~\bibnamefont {Huang}}, \bibinfo {author} {\bibfnamefont {X.}~\bibnamefont {Meng}}, \bibinfo {author} {\bibfnamefont {C.}~\bibnamefont {Zhang}}, \bibinfo {author} {\bibfnamefont {T.}~\bibnamefont {Okuda}}, \bibinfo {author} {\bibfnamefont {K.}~\bibnamefont {Shimada}}, \bibinfo {author} {\bibfnamefont
  {S.}~\bibnamefont {Cui}}, \bibinfo {author} {\bibfnamefont {Y.}~\bibnamefont {Zhao}}, \bibinfo {author} {\bibfnamefont {G.-H.}\ \bibnamefont {Cao}}, \bibinfo {author} {\bibfnamefont {S.}~\bibnamefont {Qiao}}, \bibinfo {author} {\bibfnamefont {J.}~\bibnamefont {Liu}}, \ and\ \bibinfo {author} {\bibfnamefont {C.}~\bibnamefont {Chen}},\ }\href {\doibase 10.1038/s41567-025-02864-2} {\bibfield  {journal} {\bibinfo  {journal} {Nat. Phys.}\ }\textbf {\bibinfo {volume} {21}},\ \bibinfo {pages} {760} (\bibinfo {year} {2025}{\natexlab{a}})}\BibitemShut {NoStop}%
\bibitem [{\citenamefont {Zhu}\ \emph {et~al.}(2019)\citenamefont {Zhu}, \citenamefont {Strempfer}, \citenamefont {Rao}, \citenamefont {Occhialini}, \citenamefont {Pelliciari}, \citenamefont {Choi}, \citenamefont {Kawaguchi}, \citenamefont {You}, \citenamefont {Mitchell}, \citenamefont {Shao-Horn} \emph {et~al.}}]{zhu2019anomalous}%
  \BibitemOpen
  \bibfield  {author} {\bibinfo {author} {\bibfnamefont {Z.}~\bibnamefont {Zhu}}, \bibinfo {author} {\bibfnamefont {J.}~\bibnamefont {Strempfer}}, \bibinfo {author} {\bibfnamefont {R.}~\bibnamefont {Rao}}, \bibinfo {author} {\bibfnamefont {C.}~\bibnamefont {Occhialini}}, \bibinfo {author} {\bibfnamefont {J.}~\bibnamefont {Pelliciari}}, \bibinfo {author} {\bibfnamefont {Y.}~\bibnamefont {Choi}}, \bibinfo {author} {\bibfnamefont {T.}~\bibnamefont {Kawaguchi}}, \bibinfo {author} {\bibfnamefont {H.}~\bibnamefont {You}}, \bibinfo {author} {\bibfnamefont {J.}~\bibnamefont {Mitchell}}, \bibinfo {author} {\bibfnamefont {Y.}~\bibnamefont {Shao-Horn}},  \emph {et~al.},\ }\href@noop {} {\bibfield  {journal} {\bibinfo  {journal} {Phys. Rev. Lett.}\ }\textbf {\bibinfo {volume} {122}},\ \bibinfo {pages} {017202} (\bibinfo {year} {2019})}\BibitemShut {NoStop}%
\bibitem [{\citenamefont {Feng}\ \emph {et~al.}(2022)\citenamefont {Feng}, \citenamefont {Zhou}, \citenamefont {{\v{S}}mejkal}, \citenamefont {Wu}, \citenamefont {Zhu}, \citenamefont {Guo}, \citenamefont {Gonz{\'a}lez-Hern{\'a}ndez}, \citenamefont {Wang}, \citenamefont {Yan}, \citenamefont {Qin} \emph {et~al.}}]{feng2022anomalous}%
  \BibitemOpen
  \bibfield  {author} {\bibinfo {author} {\bibfnamefont {Z.}~\bibnamefont {Feng}}, \bibinfo {author} {\bibfnamefont {X.}~\bibnamefont {Zhou}}, \bibinfo {author} {\bibfnamefont {L.}~\bibnamefont {{\v{S}}mejkal}}, \bibinfo {author} {\bibfnamefont {L.}~\bibnamefont {Wu}}, \bibinfo {author} {\bibfnamefont {Z.}~\bibnamefont {Zhu}}, \bibinfo {author} {\bibfnamefont {H.}~\bibnamefont {Guo}}, \bibinfo {author} {\bibfnamefont {R.}~\bibnamefont {Gonz{\'a}lez-Hern{\'a}ndez}}, \bibinfo {author} {\bibfnamefont {X.}~\bibnamefont {Wang}}, \bibinfo {author} {\bibfnamefont {H.}~\bibnamefont {Yan}}, \bibinfo {author} {\bibfnamefont {P.}~\bibnamefont {Qin}},  \emph {et~al.},\ }\href@noop {} {\bibfield  {journal} {\bibinfo  {journal} {Nat. Electron.}\ }\textbf {\bibinfo {volume} {5}},\ \bibinfo {pages} {735} (\bibinfo {year} {2022})}\BibitemShut {NoStop}%
\bibitem [{\citenamefont {Tschirner}\ \emph {et~al.}(2023)\citenamefont {Tschirner}, \citenamefont {Ke{\ss}ler}, \citenamefont {Gonzalez~Betancourt}, \citenamefont {Kotte}, \citenamefont {Kriegner}, \citenamefont {B{\"u}chner}, \citenamefont {Dufouleur}, \citenamefont {Kamp}, \citenamefont {Jovic}, \citenamefont {Smejkal} \emph {et~al.}}]{tschirner2023saturation}%
  \BibitemOpen
  \bibfield  {author} {\bibinfo {author} {\bibfnamefont {T.}~\bibnamefont {Tschirner}}, \bibinfo {author} {\bibfnamefont {P.}~\bibnamefont {Ke{\ss}ler}}, \bibinfo {author} {\bibfnamefont {R.~D.}\ \bibnamefont {Gonzalez~Betancourt}}, \bibinfo {author} {\bibfnamefont {T.}~\bibnamefont {Kotte}}, \bibinfo {author} {\bibfnamefont {D.}~\bibnamefont {Kriegner}}, \bibinfo {author} {\bibfnamefont {B.}~\bibnamefont {B{\"u}chner}}, \bibinfo {author} {\bibfnamefont {J.}~\bibnamefont {Dufouleur}}, \bibinfo {author} {\bibfnamefont {M.}~\bibnamefont {Kamp}}, \bibinfo {author} {\bibfnamefont {V.}~\bibnamefont {Jovic}}, \bibinfo {author} {\bibfnamefont {L.}~\bibnamefont {Smejkal}},  \emph {et~al.},\ }\href@noop {} {\bibfield  {journal} {\bibinfo  {journal} {APL Mater.}\ }\textbf {\bibinfo {volume} {11}} (\bibinfo {year} {2023})}\BibitemShut {NoStop}%
\bibitem [{\citenamefont {Bose}\ \emph {et~al.}(2022)\citenamefont {Bose}, \citenamefont {Schreiber}, \citenamefont {Jain}, \citenamefont {Shao}, \citenamefont {Nair}, \citenamefont {Sun}, \citenamefont {Zhang}, \citenamefont {Muller}, \citenamefont {Tsymbal}, \citenamefont {Schlom} \emph {et~al.}}]{bose2022tilted}%
  \BibitemOpen
  \bibfield  {author} {\bibinfo {author} {\bibfnamefont {A.}~\bibnamefont {Bose}}, \bibinfo {author} {\bibfnamefont {N.~J.}\ \bibnamefont {Schreiber}}, \bibinfo {author} {\bibfnamefont {R.}~\bibnamefont {Jain}}, \bibinfo {author} {\bibfnamefont {D.-F.}\ \bibnamefont {Shao}}, \bibinfo {author} {\bibfnamefont {H.~P.}\ \bibnamefont {Nair}}, \bibinfo {author} {\bibfnamefont {J.}~\bibnamefont {Sun}}, \bibinfo {author} {\bibfnamefont {X.~S.}\ \bibnamefont {Zhang}}, \bibinfo {author} {\bibfnamefont {D.~A.}\ \bibnamefont {Muller}}, \bibinfo {author} {\bibfnamefont {E.~Y.}\ \bibnamefont {Tsymbal}}, \bibinfo {author} {\bibfnamefont {D.~G.}\ \bibnamefont {Schlom}},  \emph {et~al.},\ }\href@noop {} {\bibfield  {journal} {\bibinfo  {journal} {Nat. Electron.}\ }\textbf {\bibinfo {volume} {5}},\ \bibinfo {pages} {267} (\bibinfo {year} {2022})}\BibitemShut {NoStop}%
\bibitem [{\citenamefont {Bai}\ \emph {et~al.}(2022)\citenamefont {Bai}, \citenamefont {Han}, \citenamefont {Feng}, \citenamefont {Zhou}, \citenamefont {Su}, \citenamefont {Wang}, \citenamefont {Liao}, \citenamefont {Zhu}, \citenamefont {Chen}, \citenamefont {Pan} \emph {et~al.}}]{bai2022observation}%
  \BibitemOpen
  \bibfield  {author} {\bibinfo {author} {\bibfnamefont {H.}~\bibnamefont {Bai}}, \bibinfo {author} {\bibfnamefont {L.}~\bibnamefont {Han}}, \bibinfo {author} {\bibfnamefont {X.}~\bibnamefont {Feng}}, \bibinfo {author} {\bibfnamefont {Y.}~\bibnamefont {Zhou}}, \bibinfo {author} {\bibfnamefont {R.}~\bibnamefont {Su}}, \bibinfo {author} {\bibfnamefont {Q.}~\bibnamefont {Wang}}, \bibinfo {author} {\bibfnamefont {L.}~\bibnamefont {Liao}}, \bibinfo {author} {\bibfnamefont {W.}~\bibnamefont {Zhu}}, \bibinfo {author} {\bibfnamefont {X.}~\bibnamefont {Chen}}, \bibinfo {author} {\bibfnamefont {F.}~\bibnamefont {Pan}},  \emph {et~al.},\ }\href@noop {} {\bibfield  {journal} {\bibinfo  {journal} {Phys. Rev. Lett.}\ }\textbf {\bibinfo {volume} {128}},\ \bibinfo {pages} {197202} (\bibinfo {year} {2022})}\BibitemShut {NoStop}%
\bibitem [{\citenamefont {Liao}\ \emph {et~al.}(2024)\citenamefont {Liao}, \citenamefont {Wang}, \citenamefont {Tien}, \citenamefont {Huang},\ and\ \citenamefont {Qu}}]{liao2024separation}%
  \BibitemOpen
  \bibfield  {author} {\bibinfo {author} {\bibfnamefont {C.-T.}\ \bibnamefont {Liao}}, \bibinfo {author} {\bibfnamefont {Y.-C.}\ \bibnamefont {Wang}}, \bibinfo {author} {\bibfnamefont {Y.-C.}\ \bibnamefont {Tien}}, \bibinfo {author} {\bibfnamefont {S.-Y.}\ \bibnamefont {Huang}}, \ and\ \bibinfo {author} {\bibfnamefont {D.}~\bibnamefont {Qu}},\ }\href@noop {} {\bibfield  {journal} {\bibinfo  {journal} {Phys. Rev. Lett.}\ }\textbf {\bibinfo {volume} {133}},\ \bibinfo {pages} {056701} (\bibinfo {year} {2024})}\BibitemShut {NoStop}%
\bibitem [{\citenamefont {Weber}\ \emph {et~al.}(2024)\citenamefont {Weber}, \citenamefont {Wust}, \citenamefont {Haag}, \citenamefont {Akashdeep}, \citenamefont {Leckron}, \citenamefont {Schmitt}, \citenamefont {Ramos}, \citenamefont {Kikkawa}, \citenamefont {Saitoh}, \citenamefont {Kl{\"a}ui} \emph {et~al.}}]{weber2024all}%
  \BibitemOpen
  \bibfield  {author} {\bibinfo {author} {\bibfnamefont {M.}~\bibnamefont {Weber}}, \bibinfo {author} {\bibfnamefont {S.}~\bibnamefont {Wust}}, \bibinfo {author} {\bibfnamefont {L.}~\bibnamefont {Haag}}, \bibinfo {author} {\bibfnamefont {A.}~\bibnamefont {Akashdeep}}, \bibinfo {author} {\bibfnamefont {K.}~\bibnamefont {Leckron}}, \bibinfo {author} {\bibfnamefont {C.}~\bibnamefont {Schmitt}}, \bibinfo {author} {\bibfnamefont {R.}~\bibnamefont {Ramos}}, \bibinfo {author} {\bibfnamefont {T.}~\bibnamefont {Kikkawa}}, \bibinfo {author} {\bibfnamefont {E.}~\bibnamefont {Saitoh}}, \bibinfo {author} {\bibfnamefont {M.}~\bibnamefont {Kl{\"a}ui}},  \emph {et~al.},\ }\href@noop {} {\bibfield  {journal} {\bibinfo  {journal} {arXiv preprint arXiv:2408.05187}\ } (\bibinfo {year} {2024})}\BibitemShut {NoStop}%
\bibitem [{\citenamefont {Fedchenko}\ \emph {et~al.}(2024)\citenamefont {Fedchenko}, \citenamefont {Min{\'a}r}, \citenamefont {Akashdeep}, \citenamefont {D’Souza}, \citenamefont {Vasilyev}, \citenamefont {Tkach}, \citenamefont {Odenbreit}, \citenamefont {Nguyen}, \citenamefont {Kutnyakhov}, \citenamefont {Wind} \emph {et~al.}}]{fedchenko2024observation}%
  \BibitemOpen
  \bibfield  {author} {\bibinfo {author} {\bibfnamefont {O.}~\bibnamefont {Fedchenko}}, \bibinfo {author} {\bibfnamefont {J.}~\bibnamefont {Min{\'a}r}}, \bibinfo {author} {\bibfnamefont {A.}~\bibnamefont {Akashdeep}}, \bibinfo {author} {\bibfnamefont {S.~W.}\ \bibnamefont {D’Souza}}, \bibinfo {author} {\bibfnamefont {D.}~\bibnamefont {Vasilyev}}, \bibinfo {author} {\bibfnamefont {O.}~\bibnamefont {Tkach}}, \bibinfo {author} {\bibfnamefont {L.}~\bibnamefont {Odenbreit}}, \bibinfo {author} {\bibfnamefont {Q.}~\bibnamefont {Nguyen}}, \bibinfo {author} {\bibfnamefont {D.}~\bibnamefont {Kutnyakhov}}, \bibinfo {author} {\bibfnamefont {N.}~\bibnamefont {Wind}},  \emph {et~al.},\ }\href@noop {} {\bibfield  {journal} {\bibinfo  {journal} {Sci. Adv.}\ }\textbf {\bibinfo {volume} {10}},\ \bibinfo {pages} {eadj4883} (\bibinfo {year} {2024})}\BibitemShut {NoStop}%
\bibitem [{\citenamefont {Hiraishi}\ \emph {et~al.}(2024)\citenamefont {Hiraishi}, \citenamefont {Okabe}, \citenamefont {Koda}, \citenamefont {Kadono}, \citenamefont {Muroi}, \citenamefont {Hirai},\ and\ \citenamefont {Hiroi}}]{hiraishi2024nonmagnetic}%
  \BibitemOpen
  \bibfield  {author} {\bibinfo {author} {\bibfnamefont {M.}~\bibnamefont {Hiraishi}}, \bibinfo {author} {\bibfnamefont {H.}~\bibnamefont {Okabe}}, \bibinfo {author} {\bibfnamefont {A.}~\bibnamefont {Koda}}, \bibinfo {author} {\bibfnamefont {R.}~\bibnamefont {Kadono}}, \bibinfo {author} {\bibfnamefont {T.}~\bibnamefont {Muroi}}, \bibinfo {author} {\bibfnamefont {D.}~\bibnamefont {Hirai}}, \ and\ \bibinfo {author} {\bibfnamefont {Z.}~\bibnamefont {Hiroi}},\ }\href@noop {} {\bibfield  {journal} {\bibinfo  {journal} {Phys. Rev. Lett.}\ }\textbf {\bibinfo {volume} {132}},\ \bibinfo {pages} {166702} (\bibinfo {year} {2024})}\BibitemShut {NoStop}%
\bibitem [{\citenamefont {Ke{\ss}ler}\ \emph {et~al.}(2024)\citenamefont {Ke{\ss}ler}, \citenamefont {Garcia-Gassull}, \citenamefont {Suter}, \citenamefont {Prokscha}, \citenamefont {Salman}, \citenamefont {Khalyavin}, \citenamefont {Manuel}, \citenamefont {Orlandi}, \citenamefont {Mazin}, \citenamefont {Valent{\'\i}} \emph {et~al.}}]{kessler2024absence}%
  \BibitemOpen
  \bibfield  {author} {\bibinfo {author} {\bibfnamefont {P.}~\bibnamefont {Ke{\ss}ler}}, \bibinfo {author} {\bibfnamefont {L.}~\bibnamefont {Garcia-Gassull}}, \bibinfo {author} {\bibfnamefont {A.}~\bibnamefont {Suter}}, \bibinfo {author} {\bibfnamefont {T.}~\bibnamefont {Prokscha}}, \bibinfo {author} {\bibfnamefont {Z.}~\bibnamefont {Salman}}, \bibinfo {author} {\bibfnamefont {D.}~\bibnamefont {Khalyavin}}, \bibinfo {author} {\bibfnamefont {P.}~\bibnamefont {Manuel}}, \bibinfo {author} {\bibfnamefont {F.}~\bibnamefont {Orlandi}}, \bibinfo {author} {\bibfnamefont {I.~I.}\ \bibnamefont {Mazin}}, \bibinfo {author} {\bibfnamefont {R.}~\bibnamefont {Valent{\'\i}}},  \emph {et~al.},\ }\href@noop {} {\bibfield  {journal} {\bibinfo  {journal} {npj Spintronics}\ }\textbf {\bibinfo {volume} {2}},\ \bibinfo {pages} {50} (\bibinfo {year} {2024})}\BibitemShut {NoStop}%
\bibitem [{\citenamefont {Occhialini}\ \emph {et~al.}(2025)\citenamefont {Occhialini}, \citenamefont {Nelson}, \citenamefont {Bombardi}, \citenamefont {Fan}, \citenamefont {Acevedo-Esteves}, \citenamefont {Comin}, \citenamefont {Basov}, \citenamefont {Musashi}, \citenamefont {Kawasaki}, \citenamefont {Uchida} \emph {et~al.}}]{occhialini2025structural}%
  \BibitemOpen
  \bibfield  {author} {\bibinfo {author} {\bibfnamefont {C.~A.}\ \bibnamefont {Occhialini}}, \bibinfo {author} {\bibfnamefont {C.}~\bibnamefont {Nelson}}, \bibinfo {author} {\bibfnamefont {A.}~\bibnamefont {Bombardi}}, \bibinfo {author} {\bibfnamefont {S.}~\bibnamefont {Fan}}, \bibinfo {author} {\bibfnamefont {R.}~\bibnamefont {Acevedo-Esteves}}, \bibinfo {author} {\bibfnamefont {R.}~\bibnamefont {Comin}}, \bibinfo {author} {\bibfnamefont {D.~N.}\ \bibnamefont {Basov}}, \bibinfo {author} {\bibfnamefont {M.}~\bibnamefont {Musashi}}, \bibinfo {author} {\bibfnamefont {M.}~\bibnamefont {Kawasaki}}, \bibinfo {author} {\bibfnamefont {M.}~\bibnamefont {Uchida}},  \emph {et~al.},\ }\href@noop {} {\bibfield  {journal} {\bibinfo  {journal} {arXiv preprint arXiv:2510.13767}\ } (\bibinfo {year} {2025})}\BibitemShut {NoStop}%
\bibitem [{\citenamefont {Kiefer}\ \emph {et~al.}(2025)\citenamefont {Kiefer}, \citenamefont {Wirth}, \citenamefont {Bertin}, \citenamefont {Becker}, \citenamefont {Bohat{\`y}}, \citenamefont {Schmalzl}, \citenamefont {Stunault}, \citenamefont {Rodr{\'\i}guez-Velamaz{\'a}n}, \citenamefont {Fabelo},\ and\ \citenamefont {Braden}}]{kiefer2025crystal}%
  \BibitemOpen
  \bibfield  {author} {\bibinfo {author} {\bibfnamefont {L.}~\bibnamefont {Kiefer}}, \bibinfo {author} {\bibfnamefont {F.}~\bibnamefont {Wirth}}, \bibinfo {author} {\bibfnamefont {A.}~\bibnamefont {Bertin}}, \bibinfo {author} {\bibfnamefont {P.}~\bibnamefont {Becker}}, \bibinfo {author} {\bibfnamefont {L.}~\bibnamefont {Bohat{\`y}}}, \bibinfo {author} {\bibfnamefont {K.}~\bibnamefont {Schmalzl}}, \bibinfo {author} {\bibfnamefont {A.}~\bibnamefont {Stunault}}, \bibinfo {author} {\bibfnamefont {J.~A.}\ \bibnamefont {Rodr{\'\i}guez-Velamaz{\'a}n}}, \bibinfo {author} {\bibfnamefont {O.}~\bibnamefont {Fabelo}}, \ and\ \bibinfo {author} {\bibfnamefont {M.}~\bibnamefont {Braden}},\ }\href@noop {} {\bibfield  {journal} {\bibinfo  {journal} {J. Phys. Condens. Matter}\ }\textbf {\bibinfo {volume} {37}},\ \bibinfo {pages} {135801} (\bibinfo {year} {2025})}\BibitemShut {NoStop}%
\bibitem [{\citenamefont {Liu}\ \emph {et~al.}(2024)\citenamefont {Liu}, \citenamefont {Zhan}, \citenamefont {Li}, \citenamefont {Liu}, \citenamefont {Cheng}, \citenamefont {Shi}, \citenamefont {Deng}, \citenamefont {Zhang}, \citenamefont {Li}, \citenamefont {Ding} \emph {et~al.}}]{liu2024absence}%
  \BibitemOpen
  \bibfield  {author} {\bibinfo {author} {\bibfnamefont {J.}~\bibnamefont {Liu}}, \bibinfo {author} {\bibfnamefont {J.}~\bibnamefont {Zhan}}, \bibinfo {author} {\bibfnamefont {T.}~\bibnamefont {Li}}, \bibinfo {author} {\bibfnamefont {J.}~\bibnamefont {Liu}}, \bibinfo {author} {\bibfnamefont {S.}~\bibnamefont {Cheng}}, \bibinfo {author} {\bibfnamefont {Y.}~\bibnamefont {Shi}}, \bibinfo {author} {\bibfnamefont {L.}~\bibnamefont {Deng}}, \bibinfo {author} {\bibfnamefont {M.}~\bibnamefont {Zhang}}, \bibinfo {author} {\bibfnamefont {C.}~\bibnamefont {Li}}, \bibinfo {author} {\bibfnamefont {J.}~\bibnamefont {Ding}},  \emph {et~al.},\ }\href@noop {} {\bibfield  {journal} {\bibinfo  {journal} {Phys. Rev. Lett.}\ }\textbf {\bibinfo {volume} {133}},\ \bibinfo {pages} {176401} (\bibinfo {year} {2024})}\BibitemShut {NoStop}%
\bibitem [{\citenamefont {Peng}\ \emph {et~al.}(2025)\citenamefont {Peng}, \citenamefont {Liu}, \citenamefont {Zhang}, \citenamefont {Zhou}, \citenamefont {Sun}, \citenamefont {Su}, \citenamefont {Wu}, \citenamefont {Zhou}, \citenamefont {Liu}, \citenamefont {Li} \emph {et~al.}}]{peng2024universal}%
  \BibitemOpen
  \bibfield  {author} {\bibinfo {author} {\bibfnamefont {X.}~\bibnamefont {Peng}}, \bibinfo {author} {\bibfnamefont {Z.}~\bibnamefont {Liu}}, \bibinfo {author} {\bibfnamefont {S.}~\bibnamefont {Zhang}}, \bibinfo {author} {\bibfnamefont {Y.}~\bibnamefont {Zhou}}, \bibinfo {author} {\bibfnamefont {Y.}~\bibnamefont {Sun}}, \bibinfo {author} {\bibfnamefont {Y.}~\bibnamefont {Su}}, \bibinfo {author} {\bibfnamefont {C.}~\bibnamefont {Wu}}, \bibinfo {author} {\bibfnamefont {T.}~\bibnamefont {Zhou}}, \bibinfo {author} {\bibfnamefont {L.}~\bibnamefont {Liu}}, \bibinfo {author} {\bibfnamefont {Y.}~\bibnamefont {Li}},  \emph {et~al.},\ }\href@noop {} {\bibfield  {journal} {\bibinfo  {journal} {Commun. Mater.}\ }\textbf {\bibinfo {volume} {6}},\ \bibinfo {pages} {177} (\bibinfo {year} {2025})}\BibitemShut {NoStop}%
\bibitem [{\citenamefont {Wenzel}\ \emph {et~al.}(2025)\citenamefont {Wenzel}, \citenamefont {Uykur}, \citenamefont {R{\"o}{\ss}ler}, \citenamefont {Schmidt}, \citenamefont {Janson}, \citenamefont {Tiwari}, \citenamefont {Dressel},\ and\ \citenamefont {Tsirlin}}]{wenzel2025fermi}%
  \BibitemOpen
  \bibfield  {author} {\bibinfo {author} {\bibfnamefont {M.}~\bibnamefont {Wenzel}}, \bibinfo {author} {\bibfnamefont {E.}~\bibnamefont {Uykur}}, \bibinfo {author} {\bibfnamefont {S.}~\bibnamefont {R{\"o}{\ss}ler}}, \bibinfo {author} {\bibfnamefont {M.}~\bibnamefont {Schmidt}}, \bibinfo {author} {\bibfnamefont {O.}~\bibnamefont {Janson}}, \bibinfo {author} {\bibfnamefont {A.}~\bibnamefont {Tiwari}}, \bibinfo {author} {\bibfnamefont {M.}~\bibnamefont {Dressel}}, \ and\ \bibinfo {author} {\bibfnamefont {A.~A.}\ \bibnamefont {Tsirlin}},\ }\href@noop {} {\bibfield  {journal} {\bibinfo  {journal} {Phys. Rev. B}\ }\textbf {\bibinfo {volume} {111}},\ \bibinfo {pages} {L041115} (\bibinfo {year} {2025})}\BibitemShut {NoStop}%
\bibitem [{\citenamefont {Smolyanyuk}\ \emph {et~al.}(2024)\citenamefont {Smolyanyuk}, \citenamefont {Mazin}, \citenamefont {Garcia-Gassull},\ and\ \citenamefont {Valent\'{\i}}}]{PhysRevB.109.134424}%
  \BibitemOpen
  \bibfield  {author} {\bibinfo {author} {\bibfnamefont {A.}~\bibnamefont {Smolyanyuk}}, \bibinfo {author} {\bibfnamefont {I.~I.}\ \bibnamefont {Mazin}}, \bibinfo {author} {\bibfnamefont {L.}~\bibnamefont {Garcia-Gassull}}, \ and\ \bibinfo {author} {\bibfnamefont {R.}~\bibnamefont {Valent\'{\i}}},\ }\href {\doibase 10.1103/PhysRevB.109.134424} {\bibfield  {journal} {\bibinfo  {journal} {Phys. Rev. B}\ }\textbf {\bibinfo {volume} {109}},\ \bibinfo {pages} {134424} (\bibinfo {year} {2024})}\BibitemShut {NoStop}%
\bibitem [{\citenamefont {Wickramaratne}\ \emph {et~al.}(2025)\citenamefont {Wickramaratne}, \citenamefont {Currie}, \citenamefont {Fields}, \citenamefont {Cress},\ and\ \citenamefont {Bennett}}]{wickramaratne2025effects}%
  \BibitemOpen
  \bibfield  {author} {\bibinfo {author} {\bibfnamefont {D.}~\bibnamefont {Wickramaratne}}, \bibinfo {author} {\bibfnamefont {M.}~\bibnamefont {Currie}}, \bibinfo {author} {\bibfnamefont {S.~S.}\ \bibnamefont {Fields}}, \bibinfo {author} {\bibfnamefont {C.~D.}\ \bibnamefont {Cress}}, \ and\ \bibinfo {author} {\bibfnamefont {S.~P.}\ \bibnamefont {Bennett}},\ }\href@noop {} {\bibfield  {journal} {\bibinfo  {journal} {arXiv preprint arXiv:2502.08872}\ } (\bibinfo {year} {2025})}\BibitemShut {NoStop}%
\bibitem [{\citenamefont {Zhang}\ \emph {et~al.}(2025{\natexlab{b}})\citenamefont {Zhang}, \citenamefont {Jeong}, \citenamefont {Buiarelli}, \citenamefont {Lee}, \citenamefont {Guo}, \citenamefont {Wen}, \citenamefont {Li}, \citenamefont {Nair}, \citenamefont {Choi}, \citenamefont {Ren}, \citenamefont {Yue}, \citenamefont {Fedorov}, \citenamefont {Mo}, \citenamefont {Kono}, \citenamefont {Lee}, \citenamefont {Low}, \citenamefont {Birol}, \citenamefont {Fernandes}, \citenamefont {Radovic}, \citenamefont {Jalan},\ and\ \citenamefont {Yi}}]{yichen2025}%
  \BibitemOpen
  \bibfield  {author} {\bibinfo {author} {\bibfnamefont {Y.}~\bibnamefont {Zhang}}, \bibinfo {author} {\bibfnamefont {S.~G.}\ \bibnamefont {Jeong}}, \bibinfo {author} {\bibfnamefont {L.}~\bibnamefont {Buiarelli}}, \bibinfo {author} {\bibfnamefont {S.}~\bibnamefont {Lee}}, \bibinfo {author} {\bibfnamefont {Y.}~\bibnamefont {Guo}}, \bibinfo {author} {\bibfnamefont {J.}~\bibnamefont {Wen}}, \bibinfo {author} {\bibfnamefont {H.}~\bibnamefont {Li}}, \bibinfo {author} {\bibfnamefont {S.}~\bibnamefont {Nair}}, \bibinfo {author} {\bibfnamefont {I.~H.}\ \bibnamefont {Choi}}, \bibinfo {author} {\bibfnamefont {Z.}~\bibnamefont {Ren}}, \bibinfo {author} {\bibfnamefont {Z.}~\bibnamefont {Yue}}, \bibinfo {author} {\bibfnamefont {A.}~\bibnamefont {Fedorov}}, \bibinfo {author} {\bibfnamefont {S.-K.}\ \bibnamefont {Mo}}, \bibinfo {author} {\bibfnamefont {J.}~\bibnamefont {Kono}}, \bibinfo {author} {\bibfnamefont {J.~S.}\ \bibnamefont {Lee}}, \bibinfo {author} {\bibfnamefont {T.}~\bibnamefont {Low}}, \bibinfo {author}
  {\bibfnamefont {T.}~\bibnamefont {Birol}}, \bibinfo {author} {\bibfnamefont {R.~M.}\ \bibnamefont {Fernandes}}, \bibinfo {author} {\bibfnamefont {M.}~\bibnamefont {Radovic}}, \bibinfo {author} {\bibfnamefont {B.}~\bibnamefont {Jalan}}, \ and\ \bibinfo {author} {\bibfnamefont {M.}~\bibnamefont {Yi}},\ }\href@noop {} {\bibfield  {journal} {\bibinfo  {journal} {arXiv preprint arXiv:2509.16361}\ } (\bibinfo {year} {2025}{\natexlab{b}})}\BibitemShut {NoStop}%
\bibitem [{\citenamefont {Jeong}\ \emph {et~al.}(2026{\natexlab{a}})\citenamefont {Jeong}, \citenamefont {Choi}, \citenamefont {Nair}, \citenamefont {Buiarelli}, \citenamefont {Pourbahari}, \citenamefont {Oh}, \citenamefont {Lin}, \citenamefont {LeBeau}, \citenamefont {Bassim}, \citenamefont {Hirai} \emph {et~al.}}]{jeong2024altermagnetic}%
  \BibitemOpen
  \bibfield  {author} {\bibinfo {author} {\bibfnamefont {S.~G.}\ \bibnamefont {Jeong}}, \bibinfo {author} {\bibfnamefont {I.~H.}\ \bibnamefont {Choi}}, \bibinfo {author} {\bibfnamefont {S.}~\bibnamefont {Nair}}, \bibinfo {author} {\bibfnamefont {L.}~\bibnamefont {Buiarelli}}, \bibinfo {author} {\bibfnamefont {B.}~\bibnamefont {Pourbahari}}, \bibinfo {author} {\bibfnamefont {J.~Y.}\ \bibnamefont {Oh}}, \bibinfo {author} {\bibfnamefont {B.~Y.}\ \bibnamefont {Lin}}, \bibinfo {author} {\bibfnamefont {J.~M.}\ \bibnamefont {LeBeau}}, \bibinfo {author} {\bibfnamefont {N.}~\bibnamefont {Bassim}}, \bibinfo {author} {\bibfnamefont {D.}~\bibnamefont {Hirai}},  \emph {et~al.},\ }\href@noop {} {\bibfield  {journal} {\bibinfo  {journal} {Proc. Natl. Acad. Sci.}\ }\textbf {\bibinfo {volume} {123}},\ \bibinfo {pages} {e2526641123} (\bibinfo {year} {2026}{\natexlab{a}})}\BibitemShut {NoStop}%
\bibitem [{\citenamefont {Jeong}\ \emph {et~al.}(2025{\natexlab{a}})\citenamefont {Jeong}, \citenamefont {Lee}, \citenamefont {Lin}, \citenamefont {Yang}, \citenamefont {Choi}, \citenamefont {Oh}, \citenamefont {Song}, \citenamefont {Lee}, \citenamefont {Nair}, \citenamefont {Choudhary} \emph {et~al.}}]{jeong2025metallicity}%
  \BibitemOpen
  \bibfield  {author} {\bibinfo {author} {\bibfnamefont {S.~G.}\ \bibnamefont {Jeong}}, \bibinfo {author} {\bibfnamefont {S.}~\bibnamefont {Lee}}, \bibinfo {author} {\bibfnamefont {B.}~\bibnamefont {Lin}}, \bibinfo {author} {\bibfnamefont {Z.}~\bibnamefont {Yang}}, \bibinfo {author} {\bibfnamefont {I.~H.}\ \bibnamefont {Choi}}, \bibinfo {author} {\bibfnamefont {J.~Y.}\ \bibnamefont {Oh}}, \bibinfo {author} {\bibfnamefont {S.}~\bibnamefont {Song}}, \bibinfo {author} {\bibfnamefont {S.~w.}\ \bibnamefont {Lee}}, \bibinfo {author} {\bibfnamefont {S.}~\bibnamefont {Nair}}, \bibinfo {author} {\bibfnamefont {R.}~\bibnamefont {Choudhary}},  \emph {et~al.},\ }\href@noop {} {\bibfield  {journal} {\bibinfo  {journal} {Proc. Natl. Acad. Sci.}\ }\textbf {\bibinfo {volume} {122}},\ \bibinfo {pages} {e2500831122} (\bibinfo {year} {2025}{\natexlab{a}})}\BibitemShut {NoStop}%
\bibitem [{\citenamefont {Berlijn}\ \emph {et~al.}(2017{\natexlab{b}})\citenamefont {Berlijn}, \citenamefont {Snijders}, \citenamefont {Delaire}, \citenamefont {Zhou}, \citenamefont {Maier}, \citenamefont {Cao}, \citenamefont {Chi}, \citenamefont {Matsuda}, \citenamefont {Wang}, \citenamefont {Koehler}, \citenamefont {Kent},\ and\ \citenamefont {Weitering}}]{PhysRevLett.118.077201}%
  \BibitemOpen
  \bibfield  {author} {\bibinfo {author} {\bibfnamefont {T.}~\bibnamefont {Berlijn}}, \bibinfo {author} {\bibfnamefont {P.~C.}\ \bibnamefont {Snijders}}, \bibinfo {author} {\bibfnamefont {O.}~\bibnamefont {Delaire}}, \bibinfo {author} {\bibfnamefont {H.-D.}\ \bibnamefont {Zhou}}, \bibinfo {author} {\bibfnamefont {T.~A.}\ \bibnamefont {Maier}}, \bibinfo {author} {\bibfnamefont {H.-B.}\ \bibnamefont {Cao}}, \bibinfo {author} {\bibfnamefont {S.-X.}\ \bibnamefont {Chi}}, \bibinfo {author} {\bibfnamefont {M.}~\bibnamefont {Matsuda}}, \bibinfo {author} {\bibfnamefont {Y.}~\bibnamefont {Wang}}, \bibinfo {author} {\bibfnamefont {M.~R.}\ \bibnamefont {Koehler}}, \bibinfo {author} {\bibfnamefont {P.~R.~C.}\ \bibnamefont {Kent}}, \ and\ \bibinfo {author} {\bibfnamefont {H.~H.}\ \bibnamefont {Weitering}},\ }\href {\doibase 10.1103/PhysRevLett.118.077201} {\bibfield  {journal} {\bibinfo  {journal} {Phys. Rev. Lett.}\ }\textbf {\bibinfo {volume} {118}},\ \bibinfo {pages} {077201} (\bibinfo {year}
  {2017}{\natexlab{b}})}\BibitemShut {NoStop}%
\bibitem [{\citenamefont {Burdett}\ \emph {et~al.}(1987)\citenamefont {Burdett}, \citenamefont {Hughbanks}, \citenamefont {Miller}, \citenamefont {Richardson},\ and\ \citenamefont {Smith}}]{doi:10.1021/ja00246a021}%
  \BibitemOpen
  \bibfield  {author} {\bibinfo {author} {\bibfnamefont {J.~K.}\ \bibnamefont {Burdett}}, \bibinfo {author} {\bibfnamefont {T.}~\bibnamefont {Hughbanks}}, \bibinfo {author} {\bibfnamefont {G.~J.}\ \bibnamefont {Miller}}, \bibinfo {author} {\bibfnamefont {J.~W.~J.}\ \bibnamefont {Richardson}}, \ and\ \bibinfo {author} {\bibfnamefont {J.~V.}\ \bibnamefont {Smith}},\ }\href {\doibase 10.1021/ja00246a021} {\bibfield  {journal} {\bibinfo  {journal} {J. Am. Chem. Soc.}\ }\textbf {\bibinfo {volume} {109}},\ \bibinfo {pages} {3639} (\bibinfo {year} {1987})}\BibitemShut {NoStop}%
\bibitem [{SI()}]{SI}%
  \BibitemOpen
  \href@noop {} {}\bibinfo {note} {See Supplemental Material at URL-will-be-inserted-by-publisher for more details on the experiment and other theoretical results.}\BibitemShut {Stop}%
\bibitem [{\citenamefont {Moriya}(1959)}]{moriya1959piezomagnetism}%
  \BibitemOpen
  \bibfield  {author} {\bibinfo {author} {\bibfnamefont {T.}~\bibnamefont {Moriya}},\ }\href@noop {} {\bibfield  {journal} {\bibinfo  {journal} {J. Phys. Chem. Solids.}\ }\textbf {\bibinfo {volume} {11}},\ \bibinfo {pages} {73} (\bibinfo {year} {1959})}\BibitemShut {NoStop}%
\bibitem [{\citenamefont {Komuro}\ \emph {et~al.}(2025)\citenamefont {Komuro}, \citenamefont {Aoyama},\ and\ \citenamefont {Ohgushi}}]{komuro2025revisiting}%
  \BibitemOpen
  \bibfield  {author} {\bibinfo {author} {\bibfnamefont {M.}~\bibnamefont {Komuro}}, \bibinfo {author} {\bibfnamefont {T.}~\bibnamefont {Aoyama}}, \ and\ \bibinfo {author} {\bibfnamefont {K.}~\bibnamefont {Ohgushi}},\ }\href@noop {} {\bibfield  {journal} {\bibinfo  {journal} {Phys. Rev. B}\ }\textbf {\bibinfo {volume} {111}},\ \bibinfo {pages} {214445} (\bibinfo {year} {2025})}\BibitemShut {NoStop}%
\bibitem [{\citenamefont {Wadehra}\ \emph {et~al.}(2025)\citenamefont {Wadehra}, \citenamefont {Z.~Gregory}, \citenamefont {Zhang}, \citenamefont {Schnitzer}, \citenamefont {Iguchi}, \citenamefont {Evan~Li}, \citenamefont {Pamuk}, \citenamefont {A.~Muller}, \citenamefont {Singer}, \citenamefont {M.~Shen},\ and\ \citenamefont {G.~Schlom}}]{https://doi.org/10.1038/s43246-025-00856-6}%
  \BibitemOpen
  \bibfield  {author} {\bibinfo {author} {\bibfnamefont {N.}~\bibnamefont {Wadehra}}, \bibinfo {author} {\bibfnamefont {B.}~\bibnamefont {Z.~Gregory}}, \bibinfo {author} {\bibfnamefont {S.}~\bibnamefont {Zhang}}, \bibinfo {author} {\bibfnamefont {N.}~\bibnamefont {Schnitzer}}, \bibinfo {author} {\bibfnamefont {Y.}~\bibnamefont {Iguchi}}, \bibinfo {author} {\bibfnamefont {Y.}~\bibnamefont {Evan~Li}}, \bibinfo {author} {\bibfnamefont {B.}~\bibnamefont {Pamuk}}, \bibinfo {author} {\bibfnamefont {D.}~\bibnamefont {A.~Muller}}, \bibinfo {author} {\bibfnamefont {A.}~\bibnamefont {Singer}}, \bibinfo {author} {\bibfnamefont {K.}~\bibnamefont {M.~Shen}}, \ and\ \bibinfo {author} {\bibfnamefont {D.}~\bibnamefont {G.~Schlom}},\ }\href@noop {} {\bibfield  {journal} {\bibinfo  {journal} {Commun. Mater.}\ }\textbf {\bibinfo {volume} {6}} (\bibinfo {year} {2025})}\BibitemShut {NoStop}%
\bibitem [{\citenamefont {Jeong}\ \emph {et~al.}(2025{\natexlab{b}})\citenamefont {Jeong}, \citenamefont {Choi}, \citenamefont {Lee}, \citenamefont {Oh}, \citenamefont {Nair}, \citenamefont {Lee}, \citenamefont {Kim}, \citenamefont {Seo}, \citenamefont {Choi}, \citenamefont {Low} \emph {et~al.}}]{jeong2025anisotropic}%
  \BibitemOpen
  \bibfield  {author} {\bibinfo {author} {\bibfnamefont {S.~G.}\ \bibnamefont {Jeong}}, \bibinfo {author} {\bibfnamefont {I.~H.}\ \bibnamefont {Choi}}, \bibinfo {author} {\bibfnamefont {S.}~\bibnamefont {Lee}}, \bibinfo {author} {\bibfnamefont {J.~Y.}\ \bibnamefont {Oh}}, \bibinfo {author} {\bibfnamefont {S.}~\bibnamefont {Nair}}, \bibinfo {author} {\bibfnamefont {J.~H.}\ \bibnamefont {Lee}}, \bibinfo {author} {\bibfnamefont {C.}~\bibnamefont {Kim}}, \bibinfo {author} {\bibfnamefont {A.}~\bibnamefont {Seo}}, \bibinfo {author} {\bibfnamefont {W.~S.}\ \bibnamefont {Choi}}, \bibinfo {author} {\bibfnamefont {T.}~\bibnamefont {Low}},  \emph {et~al.},\ }\href@noop {} {\bibfield  {journal} {\bibinfo  {journal} {Sci. Adv.}\ }\textbf {\bibinfo {volume} {11}},\ \bibinfo {pages} {eadw7125} (\bibinfo {year} {2025}{\natexlab{b}})}\BibitemShut {NoStop}%
\bibitem [{\citenamefont {Jeong}\ \emph {et~al.}(2026{\natexlab{b}})\citenamefont {Jeong}, \citenamefont {Lin}, \citenamefont {Jin}, \citenamefont {Choi}, \citenamefont {Lee}, \citenamefont {Yang}, \citenamefont {Nair}, \citenamefont {Choudhary}, \citenamefont {Parikh}, \citenamefont {Santhosh} \emph {et~al.}}]{jeong2025strainstabilizedinterfacialpolarizationtunes}%
  \BibitemOpen
  \bibfield  {author} {\bibinfo {author} {\bibfnamefont {S.~G.}\ \bibnamefont {Jeong}}, \bibinfo {author} {\bibfnamefont {B.~Y.}\ \bibnamefont {Lin}}, \bibinfo {author} {\bibfnamefont {M.}~\bibnamefont {Jin}}, \bibinfo {author} {\bibfnamefont {I.~H.}\ \bibnamefont {Choi}}, \bibinfo {author} {\bibfnamefont {S.}~\bibnamefont {Lee}}, \bibinfo {author} {\bibfnamefont {Z.}~\bibnamefont {Yang}}, \bibinfo {author} {\bibfnamefont {S.}~\bibnamefont {Nair}}, \bibinfo {author} {\bibfnamefont {R.}~\bibnamefont {Choudhary}}, \bibinfo {author} {\bibfnamefont {J.}~\bibnamefont {Parikh}}, \bibinfo {author} {\bibfnamefont {A.}~\bibnamefont {Santhosh}},  \emph {et~al.},\ }\href@noop {} {\bibfield  {journal} {\bibinfo  {journal} {Nat. Commun.}\ } (\bibinfo {year} {2026}{\natexlab{b}})}\BibitemShut {NoStop}%
\bibitem [{\citenamefont {Stoeberl}\ \emph {et~al.}(2020)\citenamefont {Stoeberl}, \citenamefont {Guedes}, \citenamefont {Abud}, \citenamefont {Jardim}, \citenamefont {Abbate},\ and\ \citenamefont {Mossanek}}]{Stoeberl_2020}%
  \BibitemOpen
  \bibfield  {author} {\bibinfo {author} {\bibfnamefont {V.}~\bibnamefont {Stoeberl}}, \bibinfo {author} {\bibfnamefont {E.~B.}\ \bibnamefont {Guedes}}, \bibinfo {author} {\bibfnamefont {F.}~\bibnamefont {Abud}}, \bibinfo {author} {\bibfnamefont {R.~F.}\ \bibnamefont {Jardim}}, \bibinfo {author} {\bibfnamefont {M.}~\bibnamefont {Abbate}}, \ and\ \bibinfo {author} {\bibfnamefont {R.~J.~O.}\ \bibnamefont {Mossanek}},\ }\href {\doibase 10.1209/0295-5075/132/47004} {\bibfield  {journal} {\bibinfo  {journal} {Europhys. Lett.}\ }\textbf {\bibinfo {volume} {132}},\ \bibinfo {pages} {47004} (\bibinfo {year} {2020})}\BibitemShut {NoStop}%
\bibitem [{\citenamefont {Julliere}(1975)}]{julliere1975tunneling}%
  \BibitemOpen
  \bibfield  {author} {\bibinfo {author} {\bibfnamefont {M.}~\bibnamefont {Julliere}},\ }\href@noop {} {\bibfield  {journal} {\bibinfo  {journal} {Phys. Lett. A}\ }\textbf {\bibinfo {volume} {54}},\ \bibinfo {pages} {225} (\bibinfo {year} {1975})}\BibitemShut {NoStop}%
\bibitem [{\citenamefont {Jiang}\ \emph {et~al.}(2023)\citenamefont {Jiang}, \citenamefont {Wang}, \citenamefont {Samanta}, \citenamefont {Zhang}, \citenamefont {Xiao}, \citenamefont {Lu}, \citenamefont {Sun}, \citenamefont {Tsymbal},\ and\ \citenamefont {Shao}}]{jiang2023prediction}%
  \BibitemOpen
  \bibfield  {author} {\bibinfo {author} {\bibfnamefont {Y.-Y.}\ \bibnamefont {Jiang}}, \bibinfo {author} {\bibfnamefont {Z.-A.}\ \bibnamefont {Wang}}, \bibinfo {author} {\bibfnamefont {K.}~\bibnamefont {Samanta}}, \bibinfo {author} {\bibfnamefont {S.-H.}\ \bibnamefont {Zhang}}, \bibinfo {author} {\bibfnamefont {R.-C.}\ \bibnamefont {Xiao}}, \bibinfo {author} {\bibfnamefont {W.}~\bibnamefont {Lu}}, \bibinfo {author} {\bibfnamefont {Y.}~\bibnamefont {Sun}}, \bibinfo {author} {\bibfnamefont {E.~Y.}\ \bibnamefont {Tsymbal}}, \ and\ \bibinfo {author} {\bibfnamefont {D.-F.}\ \bibnamefont {Shao}},\ }\href@noop {} {\bibfield  {journal} {\bibinfo  {journal} {Phys. Rev. B}\ }\textbf {\bibinfo {volume} {108}},\ \bibinfo {pages} {174439} (\bibinfo {year} {2023})}\BibitemShut {NoStop}%
\bibitem [{\citenamefont {Gurung}\ \emph {et~al.}(2024)\citenamefont {Gurung}, \citenamefont {Elekhtiar}, \citenamefont {Luo}, \citenamefont {Shao},\ and\ \citenamefont {Tsymbal}}]{gurung2024nearly}%
  \BibitemOpen
  \bibfield  {author} {\bibinfo {author} {\bibfnamefont {G.}~\bibnamefont {Gurung}}, \bibinfo {author} {\bibfnamefont {M.}~\bibnamefont {Elekhtiar}}, \bibinfo {author} {\bibfnamefont {Q.-Q.}\ \bibnamefont {Luo}}, \bibinfo {author} {\bibfnamefont {D.-F.}\ \bibnamefont {Shao}}, \ and\ \bibinfo {author} {\bibfnamefont {E.~Y.}\ \bibnamefont {Tsymbal}},\ }\href@noop {} {\bibfield  {journal} {\bibinfo  {journal} {Nat. Commun.}\ }\textbf {\bibinfo {volume} {15}},\ \bibinfo {pages} {10242} (\bibinfo {year} {2024})}\BibitemShut {NoStop}%
\end{thebibliography}
\end{document}